\documentclass[twocolumn,
                aps,
                superscriptaddress,
                prb, 
                notitlepage,
                10pt,
		]{revtex4-2}

\usepackage{amssymb}
\usepackage{amsmath}
\usepackage{dsfont}
\usepackage{float}
\usepackage{graphicx}
\usepackage{dcolumn}
\usepackage{tabularx}
\newcolumntype{C}{>{\centering\arraybackslash}X}
\newcolumntype{L}{>{\raggedright\arraybackslash}X} 
\usepackage{url}
\usepackage[colorlinks=true,breaklinks=true,allcolors=blue]{hyperref}
\usepackage{color}
\usepackage{xcolor}
\usepackage{bbold}
\usepackage[noabbrev]{cleveref}
\Crefname{figure}{Fig.}{Figs.}
\Crefname{equation}{Eq.}{Eqs.}
\Crefname{table}{Tab.}{Tabs.}

\makeatletter
\let\cat@comma@active\@empty
\makeatother

\begin{document}

\title{Leveraging recurrence in neural network wavefunctions for\newline
large-scale simulations of Heisenberg antiferromagnets
on\newline
the triangular lattice}

\author{M. Schuyler Moss}
\email{msmoss@uwaterloo.ca}
\affiliation{Department of Physics and Astronomy, University of Waterloo, Ontario, N2L 3G1, Canada}
\affiliation{Perimeter Institute for Theoretical Physics, Waterloo, Ontario, N2L 2Y5, Canada}

\author{Roeland Wiersema}
\affiliation{Center for Computational Quantum Physics, Flatiron Institute, 162 Fifth Avenue, New York, NY 10010, USA}

\author{Mohamed Hibat-Allah}
\affiliation{Department of Applied Mathematics, University of Waterloo, Waterloo, ON N2L 3G1, Canada}
\affiliation{Vector Institute,  Toronto,  Ontario,  M5G 0C6,  Canada}

 \author{Juan Carrasquilla} %\orcidlink{0000-0001-7263-3462}
\affiliation{Institute for Theoretical Physics, ETH Zürich, 8093, Switzerland}

\author{Roger G. Melko}
\affiliation{Department of Physics and Astronomy, University of Waterloo, Ontario, N2L 3G1, Canada}
\affiliation{Perimeter Institute for Theoretical Physics, Waterloo, Ontario, N2L 2Y5, Canada}

\begin{abstract}

Variational Monte Carlo simulations have been crucial for understanding quantum many-body systems, especially when the Hamiltonian is frustrated and the ground-state wavefunction has a non-trivial sign structure.
In this paper, we use recurrent neural network (RNN) wavefunction ans\"{a}tze to study the triangular-lattice antiferromagnetic Heisenberg model (TLAHM) for lattice sizes up to $30\times30$.
In a recent study [M. S. Moss \emph{et al.} arXiv:2502.17144], the authors demonstrated how RNN wavefunctions can be iteratively retrained in order to obtain variational results for multiple lattice sizes with a reasonable amount of compute.
That study, which looked at the sign-free, square-lattice antiferromagnetic Heisenberg model, showed favorable scaling properties, allowing accurate finite-size extrapolations to the thermodynamic limit.
In contrast, our present results illustrate in detail the relative difficulty in simulating the sign-problematic TLAHM.
We find that the accuracy of our simulations can be significantly improved by transforming the Hamiltonian with a judicious choice of basis rotation.
We also show that a similar benefit can be achieved by using variational neural annealing, an alternative optimization technique that minimizes a pseudo free energy.
Ultimately, we are able to obtain estimates of the ground-state properties of the TLAHM in the thermodynamic limit that are in close agreement with values in the literature, showing that RNN wavefunctions provide a powerful toolbox for performing finite-size scaling studies for frustrated quantum many-body systems.

\end{abstract}

\maketitle

\section{Introduction}

The triangular-lattice spin-$\frac{1}{2}$ antiferromagnetic Heisenberg model (TLAHM) is one of the standard examples of frustrated quantum magnetism. While it is well-known today that the ground state of the TLAHM exhibits 120$^\circ$ magnetic order, it was originally believed that the frustration in the lattice would lead to a magnetically disordered ground state~\cite{anderson_resonating_1973}. For a long time, there was strong disagreement about the nature of this Hamiltonian's ground state, which illustrates the challenge this problem presents. 

Unlike for its square-lattice counterpart, quantum Monte Carlo (QMC) simulations of the TLAHM suffer from a sign problem. As such, early numerical results for the ground state of the TLAHM were mostly limited to exact diagonalization studies on small system sizes~\cite{marland_frustration_1979,oguchi_ground-state_1986,nishimori_ground_1988,bernu_signature_1992,leung_spin-12_1993,bernu_exact_1994} or to approximate variational calculations~\cite{fazekas_ground_1974,miyashita_variational_1984,kalmeyer_equivalence_1987,huse_simple_1988}. Both of these approaches, however, have consequential weaknesses.
One can only perform exact diagonalization for relatively small system sizes, where strong finite-size effects can severely bias extrapolations to the thermodynamic limit.
On the other hand, one can simulate much larger systems using variational Monte Carlo (VMC), but the accuracy of any variational simulation is strongly dependent on the choice of ans\"{a}tze. 
Some of these early numerical results supported the original idea~\cite{anderson_resonating_1973} that the quantum fluctuations in the TLAHM would lead to a disordered ground state~\cite{oguchi_ground-state_1986,nishimori_ground_1988,leung_spin-12_1993}, while other studies concluded the exact opposite, namely, a ground state with long-range antiferromagnetic order~\cite{miyashita_variational_1984,huse_simple_1988,bernu_signature_1992,bernu_exact_1994}.
Even though semi-classical spin-wave theory predicted an ordered ground state in the thermodynamic limit~\cite{nishimori_ground-state_1985,jolicoeur_spin-wave_1989,j_miyake_spin-wave_1992, chubukov_large-s_1994},
the lack of consensus among the numerical results left room for doubt.

Eventually, new techniques~\cite{sorella_green_1998} allowed for QMC simulations of sign-problematic ground states with controlled errors. Green's function Monte Carlo with stochastic reconfiguration was successfully applied to the TLAHM~\cite{capriotti_long-range_1999}, and these simulations strongly supported the picture of a ground state with long-range antiferromagnetic order. These QMC results, alongside the direct observation of properties of the low-lying energy spectrum~\cite{bernu_signature_1992,bernu_exact_1994} that supported the assumptions and thus the conclusions of spin-wave theory~\cite{jolicoeur_spin-wave_1989,j_miyake_spin-wave_1992, chubukov_large-s_1994}, led to the eventual consensus that the TLAHM hosts an antiferromagnetically ordered ground state with spins that form $120^\circ$ angles relative to their nearest neighbors.

Over time, the availability of computational resources has increased dramatically, allowing for numerical simulations of a much larger scale. While many types of numerical methods improved with the increasing compute power~\cite{singh_three-sublattice_1992,singh1999,zheng_excitation_2006}, variational methods, in particular, benefited  from the ability to employ far more expressive ans\"{a}tze with ever-increasing numbers of variational parameters. 
For instance, the density matrix renormalization group (DMRG)~\cite{white_density_1992} has been successfully applied to two-dimensional tensor networks~\cite{xiang_two-dimensional_2001} and 
matrix product states defined on cylinders~\cite{white_neel_2007,huang_magnetization_2024} to extract accurate estimates of the ground-state properties of the TLAHM. The additional compute also allowed for traditional VMC simulations with variational ans\"{a}tze which have many parameters and are therefore more expressive~\cite{weber_magnetism_2006,heidarian_spin-frac12_2009,mezzacapo_ground-state_2010,kaneko_gapless_2021,iqbal_spin_2016,hasik_incommensurate_2024}. 
In recent years, artificial neural networks have emerged as promising candidates for ans\"{a}tze~\cite{carleo_solving_2017}. These so-called neural quantum states (NQS) have led to many state-of-the-art results~\cite{roth_high-accuracy_2023,viteritti_transformer_2023,sprague2024,chen_empowering_2024,rende_simple_2024,smith_ground_2024}; however, there remain open questions regarding the limits of NQS.
For example, frustration and/or a non-trivial sign structure in the ground-state wavefunction have been presented as potential problems for the optimization of 
NQS~\cite{westerhout2020generalization,szabo2020neural,bukov2021learning}.
Even though the architectures typically used as NQS are highly expressive~\cite{kolmogorov1957representation,hornik1991approximation,schafer2006recurrent,le2008representational,zhou2020universality}, a challenging optimization task could still make it difficult to obtain accurate variational results, especially for large system sizes. As mentioned, without \emph{accurate} variational results for multiple \emph{sufficiently-large} system sizes, one cannot perform reliable finite-size scaling studies. Indeed, finite-size scaling studies are rare in the field of NQS.

The TLAHM is therefore an important proving ground for NQS simulations~\cite{ferrari_neural_2019,hibat-allah_supplementing_2024,kochkov_learning_2021,fu_lattice_2022,wu_tensor-network_2023,roth_high-accuracy_2023}
because it is a frustrated Hamiltonian and its ground state has a non-trivial sign structure, but the nature of the ground state is theoretically well-understood.
This work focuses on benchmarking the performance of two-dimensional recurrent neural network (RNN) wavefunctions on the TLAHM. 
For small system sizes, we investigate how different basis rotations of the Hamiltonian affect the ground state search, and we utilize a technique called variational neural annealing to overcome the frustration in the Hamiltonian.
Then, we iteratively retrain our most accurate wavefunctions, and demonstrate the ability of RNNs to scale and generalize to large system sizes. Using these results, we perform a finite-size scaling study and obtain an estimate of the ground-state energy in the limit $N\rightarrow\infty$ which is in close agreement with other values found in the literature. Furthermore, we confirm the existence of the $120^\circ$ N\'eel-ordered ground state with a finite-valued estimate of the sublattice magnetization in the thermodynamic limit. Having previously shown the success of these methods for the square-lattice Heisenberg antiferromagnet~\cite{moss_leveraging_2025}, this work demonstrates that they are also effective for frustrated Hamiltonians with ground states that have a non-trivial sign structure.  

\section{Methods}
\label{sec:methods}

We now describe our neural network architecture and optimization strategy.  
This work is closely related to our recent studies on the square-lattice antiferromagnetic Heisenberg model~\cite{moss_leveraging_2025} and here we focus only on the adaptations necessary for this work. For more details, we refer the reader to Ref.~\cite{moss_leveraging_2025}. In  \Cref{app:training_details} we provide an in-depth description of the training procedure used to obtain the results presented in the main text, including information on all relevant hyperparameters.

\subsection{The Heisenberg antiferromagnet}
\label{subsec:TLAHM}

The TLAHM is defined as, 
\begin{equation}
    \hat{H} = \sum_{\langle i j \rangle} \vec{S}_i\cdot\vec{S}_j,
    \label{eq:hamiltonian}
\end{equation}
where $\langle i,j\rangle$ are the nearest-neighbor interactions on an $L\times L$ triangular lattice with $N=L^2$ spin-$\frac{1}{2}$ spins. Note that $\hat{H}$ obeys SU(2) symmetry. 

The energy of a classical antiferromagnet is minimized when all spins are anti-aligned with their nearest neighbors. However, on a triangular lattice, it is impossible to anti-align all three spins on a single triangular plaquette. This phenomenon is called \emph{geometric frustration}, which, for a classical antiferromagnet on the triangular lattice, is known to lead to a highly-degenerate energy spectrum and a disordered ground state~\cite{wannier_antiferromagnetism_1950}. 
The disorder in the ground state of the classical antiferromagnet partially inspired Anderson's resonating valence bond (RVB) theory--his proposal for the description of a magnetically disordered ground state for the TLAHM~\cite{anderson_resonating_1973}. This proposal initiated the still-ongoing pursuit of quantum spin liquids~\cite{balents_spin_2010}.

As it turns out, the true nature of the ground state of the TLAHM is an antiferromagnetically ordered ground state characterized by a finite sublattice magnetization in the thermodynamic limit. 
The existence of long-range antiferromagnetic order means that the SU(2) symmetry of the Hamiltonian is spontaneously broken in the thermodynamic limit, since any SU(2)-symmetric Hamiltonian only has SU(2)-symmetric eigenstates and SU(2)-symmetric states will exhibit a sublattice magnetization of zero~\cite{bernu_signature_1992,bernu_exact_1994}. 
Spontaneous symmetry breaking (SSB) of a continuous symmetry manifests itself in finite-size systems as a set of low-lying energy eigenstates $\vert 0\rangle$ known as Anderson's tower of states~\cite{anderson_approximate_1952}.
The states in $\vert 0\rangle$ can be combined into a superposition that localizes the direction of the sublattice magnetization and breaks global spin rotation symmetry, i.e., the SU(2) symmetry~\cite{anderson_approximate_1952}. The tower of states becomes degenerate with the ground state as $\sim 1/N$, such that in the thermodynamic limit, $N\rightarrow\infty$, the ground state is this symmetry-broken superposition. 
Notably, the spin-wave theory treatment of spin$-\frac{1}{2}$ systems begins by assuming the existence of such a tower of states~\cite{anderson_approximate_1952}.
Therefore, the numerical observation of the tower of states for the TLAHM~\cite{bernu_signature_1992,bernu_exact_1994} helped to validate the predictions of an ordered ground state that came from spin-wave theory~\cite{nishimori_ground-state_1985,jolicoeur_spin-wave_1989,j_miyake_spin-wave_1992}.

\subsection{Transforming the basis of the Hamiltonian with local unitaries}
\label{subsec:signrules}

Importantly, the triangular lattice is non-bipartite. Marshall, based on the work of Peierls~\cite{peierls}, introduced a way to transform the Heisenberg Hamiltonian, defined on a bipartite lattice, with a local unitary, $\mathcal{U}_\text{sq}$, such that the transformed Hamiltonian has non-positive off-diagonal elements~\cite{marshall_antiferromagnetism_1955}. This result is often referred to as the Marshall-Peierls Sign Rule (MPSR). Hamiltonians with non-positive off-diagonal elements are called stoquastic, and it is known that their ground-state wavefunction will have only non-negative amplitudes~\cite{perron_zur_1907,frobenius_uber_1912}. It is worth noting that the locality of the unitary transformation preserves the locality of the original Hamiltonian, which has important implications for the computational cost of VMC simulations.
Unfortunately, these results do not extend to non-bipartite lattices. 
For the triangular lattice, there is no known local unitary that can transform \Cref{eq:hamiltonian} into a stoquastic Hamiltonian. The presence of positive off-diagonal elements in the Hamiltonian gives rise to the infamous negative sign problem in QMC. The presence of positive off-diagonal elements in a Hamiltonian also means that the ground-state wavefunction could have negative amplitudes as well as positive ones, i.e., a non-trivial sign structure. 

Even though there is no formal sign problem for NQS, it has been shown that local unitary transformations that reduce the average sign of the ground-state wavefunction can stabilize the optimization~\cite{choo_two-dimensional_2019}.
For the TLAHM, one such approach is to transform the original Hamiltonian according to the unitary given by the MPSR, $\mathcal{U}_\text{sq}$. To do this, one must first define two sublattices A$_{\text{sq}}$ and B$_{\text{sq}}$ such that spins on one sublattice only interact with spins on the opposite sublattice for a square lattice with nearest neighbor couplings. Each sublattice has $N/2$ spins if $N$ is even. See \Cref{app:sign_rules} for a depiction of these sublattices. This unitary is defined as 
\begin{align}
    \mathcal{U}_\text{sq} = \text{exp}\left( -\text{i}\pi\sum_{j \in \text{B}_{\text{sq}}} \hat{S}^z_j \right),
    \label{eq:sq_sign_rule}
\end{align}
which rotates all spins on sublattice B$_{\text{sq}}$ by $\pi$ around the $z$-axis.

Rather than trying to make the Hamiltonian approximately stoquastic, one can use knowledge of the ground state to transform the Hamiltonian. 
The ground state of the TLAHM is known to display three-sublattice 120$^{\circ}$ magnetic order in the ground state.
As such, it is common to use the ``120$^{\circ}$ transformation''~\cite{miyashita_variational_1984,capriotti_quantum_2001} which is defined as follows,
\begin{align}
    \mathcal{U}_\text{tri} = \text{exp}\left( -\frac{2\pi\text{i}}{3}\left[ \sum_{b\in\text{B}_\text{tri}} \hat{S}^z_b - \sum_{c\in\text{C}_\text{tri}} \hat{S}^z_c\right] \right),
    \label{eq:tri_sign_rule}
\end{align}
where spins on sublattice B$_{\text{tri}}$ are rotated by $-\frac{2\pi}{3}$ around the $z$ axis and spins on sublattice C$_{\text{tri}}$ are rotated by $+\frac{2\pi}{3}$ around the $z$ axis. 
The three sublattices are defined such that the nearest neighbors on a triangular lattice belong to different sublattices. These sublattices are also shown in \Cref{app:sign_rules}. 

The effect of both unitary transformations on the TLAHM Hamiltonian defined in \Cref{eq:hamiltonian} is discussed in more detail in \Cref{app:sign_rules}. 

\subsection{Variational Optimization}
\label{subsec:vmc}

Variational Monte Carlo (VMC) is one of the standard tools for simulating quantum many-body ground states. VMC simulations involve defining an appropriate variational ansatz $\vert\Psi_\mathcal{W}\rangle$ (with trainable parameters $\mathcal{W}$) that is capable of capturing the important features of the target ground state~\cite{becca_quantum_2017}.
We can calculate the variational energy for a given Hamiltonian $\hat{H}$ as
\begin{align}
    E_\mathcal{W} &\equiv \frac{\langle \Psi_\mathcal{W}\vert \hat{H}\vert \Psi_\mathcal{W}\rangle}{\langle \Psi_\mathcal{W}\vert \Psi_\mathcal{W}\rangle}\nonumber
    \\
    & \approx \frac{1}{N_s}\sum_{\vec{\sigma}\sim p_\mathcal{W}(\vec{\sigma})} H_\text{loc}(\vec{\sigma}),
    \label{eq:variationalenergy}
\end{align}
where we sample $N_s$ spin configurations $\vec{\sigma}\in\{0,1\}^N$ from $p_\mathcal{W} \equiv |\Psi_\mathcal{W}|^2$. For local Hamiltonians, the local energy,
\begin{align}
    H_\text{loc}(\vec{\sigma}) \equiv \frac{\langle \vec{\sigma}\vert\hat{H}\vert\Psi_\mathcal{W}\rangle}{\langle \vec{\sigma}\vert\Psi_\mathcal{W}\rangle},
    \label{eq:localenergies}
\end{align}
can be efficiently estimated from samples~\cite{becca_quantum_2017}.
The variational principle guarantees that the variational energy $E_\mathcal{W}$ is an upper bound on the true ground-state energy. Therefore, we can minimize the variational energy defined by \Cref{eq:variationalenergy} in order to obtain an optimized wavefunction $\vert\Psi_\mathcal{W}\rangle$ that is hopefully close to the true ground state of $\hat{H}$. 

When dealing with challenging optimization landscapes with many local minima, it is common to perform variational neural annealing (VNA)~\cite{roth_iterative_2020,hibat-allah_variational_2021,roth_high-accuracy_2023,hibat-allah_investigating_2023,hibat-allah_supplementing_2024}. Instead of minimizing the variational energy $E_\mathcal{W}$ alone, one can compute a classical pseudo-entropy in order to define a variational free energy,
\begin{align}
    F_\mathcal{W}(t) = E_\mathcal{W} - T(t)S_\text{classical}(p_\mathcal{W}),
    \label{eq:freelocs}
\end{align}
where $S_\text{classical} = \frac{1}{N_s}\sum_{\vec{\sigma}\sim p_\mathcal{W}(\vec{\sigma})}\log p_\mathcal{W}(\vec{\sigma})$ is an estimator for the Shannon entropy of the probability distribution encoded by the variational ansatz $p_\mathcal{W}(\vec{\sigma})$. The quantity $T(t)$ is a pseudo-temperature that is annealed from some initial temperature $T_0$ down to zero as a function of the training step $t$. As with the variational energy, the variational free energy can be minimized with an appropriately chosen optimization scheme. 
In this work, we use the Adam optimizer~\cite{kingma2017adammethodstochasticoptimization} to minimize $F_\mathcal{W}(t)$ and $E_\mathcal{W}$.

When optimizing a variational wavefunction using VNA, there are possibly two distinct phases of the training.
If the initial temperature $T_0$ is large enough, the entropy term will dominate the variational free energy defined in \Cref{eq:freelocs} such that the loss will be minimized when the entropy is maximized. During this phase of the training, the distribution encoded by the variational ansatz $p_\mathcal{W}(\vec{\sigma})$ will become more uniform. In other words, spin configurations sampled from $p_\mathcal{W}(\vec{\sigma})$ will span the Hilbert space more broadly than they would if the ansatz were trained using traditional variational optimization. 
This regime corresponds to the first phase of training, which can be thought of as an ``exploration phase''. 
As the temperature is lowered during the annealing process, the energy term will begin to dominate, and high probabilities will be assigned to spin configurations that lower the variational energy. This regime corresponds to the second phase of training, which can be considered an ``exploitation phase''.

The estimation of the variational energy (and the pseudo-entropy) occurs at every step of the optimization. Therefore, the ease with which one can compute these quantities plays an important role in a VMC simulation. The cost of computing the local energies defined by \Cref{eq:localenergies} scales with the number of off-diagonal terms in the Hamiltonian because the variational ansatz must be evaluated for each sample $\vec{\sigma}^{\prime}$ connected to the original sample $\vec{\sigma}$ by $\hat{H}$~\cite{schmitt2020quantum}.
For the TLAHM the number of off-diagonal terms is equal to the number of nearest-neighbor interactions, which scales as $\mathcal{O}(N)$. If a basis transformation is applied to the original Hamiltonian $\hat{H}$, it is desirable that the locality of $\hat{H}$ is preserved so that the computational cost of computing the local energy does not increase.

\subsection{Recurrent neural network wavefunctions}
\label{methods:rnns}

In this work, we employ two-dimensional (2D) recurrent neural networks as our variational ans\"{a}tze~\cite{hibat-allah_recurrent_2020}. The RNN is an autoregressive neural network that models a joint probability distribution as a product of conditional probability distributions according to the chain rule of probabilities,
\begin{align}
    p(\vec{\sigma}) = p(\sigma_1)p(\sigma_2\vert\sigma_1)\dots p(\sigma_N\vert\sigma_{N-1},\dots,\sigma_2,\sigma_1)
    \label{eq:chain_rule_of_probabilities}.
\end{align}
As a wavefunction, the joint probability distribution is the distribution over a full spin configuration $\vec{\sigma} = (\sigma_1,\sigma_2,\dots,\sigma_N)$ and the conditional probabilities are over individual spins in the configuration $\sigma_i$. Using this construction, independent samples for each spin $\sigma_i$ can be obtained directly from the corresponding conditional probability $p(\sigma_i\vert\sigma_{j<i})$. Samples drawn from the conditional probabilities are independent, which is desirable for VMC simulations where sampling plays an important role.

\begin{figure}
    \centering
    \includegraphics[width=0.35\textwidth]{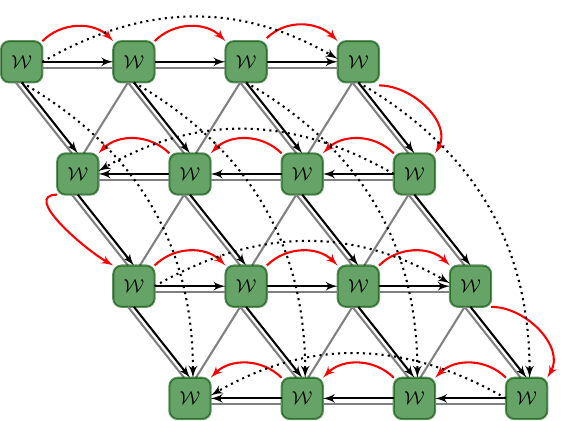}
    \caption{A 2D RNN wavefunction defined for a triangular lattice with $L = 4$. The bonds of the $4\times4$ triangular lattice are shown in grey to illustrate how the RNN structure maps to the underlying lattice. The autoregressive sequence is defined by the red arrows. Sampling and inference are performed along this path. The information in the network, stored in the hidden vectors, is passed in two directions along the black arrows. Notably, the black arrows follow the nearest-neighbor interactions of a square lattice. The black dotted arrows show how pseudo-periodic boundary connections can be built into the RNN wavefunction. Both the two-dimensional information passing and the pseudo-periodic boundary connections are implemented in a causal way such that the autoregressive sequence is not violated.}
    \label{fig:RNN}
\end{figure}

\Cref{fig:RNN} illustrates the structure of a 2D RNN wavefunction. As a consequence of the autoregressive construction, sampling from the conditional probabilities and evaluating the wavefunction on input samples (inference) are performed one spin at a time along the \emph{autoregressive sequence}, a chosen one-dimensional path through the network. How this path is defined can have important consequences during the training of an autoregressive model~\cite{teoh_autoregressive_2024}. The autoregressive sequence that we employ is highlighted with red arrows. Information in the network is stored in \emph{hidden vectors} $\vec{h}$. These hidden vectors can pass through the network in any way, so long as the autoregressive sequence is  not violated. In the original formulations of RNNs, the hidden vectors were passed along the one-dimensional autoregressive path~\cite{lipton_critical_2015}, but we employ the 2D RNN~\cite{hibat-allah_recurrent_2020,moss_enhancing_2023,moss_leveraging_2025} which is advantageous for simulating 2D physical systems. We also include pseudo-periodic boundary connections in the RNN when simulating physical systems with periodic boundary conditions~\cite{luo_gauge-invariant_2023,hibat-allah_investigating_2023,hibat-allah_recurrent_2024,moss_leveraging_2025}.

The hidden vectors are processed and computed by the main building blocks of an RNN: the \emph{recurrent cells} of the network, which are represented by the green boxes in \Cref{fig:RNN}. These cells take in a set of hidden vectors $\vec{h}_\text{input}$, each of a chosen size $d_h$, and a set of single spin states $\sigma_\text{input}$ and output a new hidden vector $\vec{h}_i$. For more details about the recurrent cells employed in this work, we refer the reader to \Cref{app:RNN}. The output hidden vector is used to compute the conditional probability over $\sigma_i$ using a dense layer with a softmax activation function:
\begin{align}
    p_\mathcal{W} (\sigma_{i}\vert\sigma_{j<i}) = \text{softmax}(\mathbf{U}\vec{h}_{i} + \vec{b}) \cdot \vec{\sigma}_{i},
\end{align}
where $\vec{\sigma}_i$ is a one-hot representation of the state of $\sigma_i$ which is sampled from the normalized distribution $[p_\mathcal{W}(\sigma_i=\uparrow\vert\sigma_{j<i}),p_\mathcal{W}(\sigma_i=\downarrow\vert\sigma_{j<i})]$. The probability for a full spin configuration $\vec{\sigma}$ is then given by $p_{\mathcal{W}}(\vec{\sigma}) = \prod_i p_\mathcal{W}(\sigma_i\vert\sigma_{j<i})$.

As mentioned, the ground state of the TLAHM has a non-trivial sign structure, so we include complex phases into our variational ansatz. 
We calculate a set of ``conditional phases'' that correspond to each spin $\sigma_i$ in the full spin configuration $\vec{\sigma}$~\cite{hibat-allah_recurrent_2020,wu_tensor-network_2023}. Similar to the probabilities, these phases are computed from the hidden vector $\vec{h}_i$ using a dense layer, but with a softsign activation function instead of a softmax:
\begin{align}
    \phi_\mathcal{W} (\sigma_{i}\vert\sigma_{j<i}) = \pi\,\text{softsign}(\mathbf{V}\vec{h}_{i} + \vec{c}) \cdot \vec{\sigma}_{i},
\end{align}
where $\vec{\sigma}_i$ is again a one-hot representation of the state of $\sigma_i$ and the softsign activation function is defined as
\begin{align}
    \text{softsign}(x) = \frac{x}{1+\vert x\vert}.
\end{align}
The phase of the wavefunction amplitude for the full spin configuration $\vec{\sigma}$ is then given by $\phi_\mathcal{W}(\vec{\sigma}) = \sum_i\phi_\mathcal{W}(\sigma_i\vert\sigma_{j<i})$. Finally, we combine the probability $p_{\mathcal{W}}(\vec\sigma)$ and the phase $\phi_\mathcal{W}(\vec{\sigma})$ to form the complex amplitude
\begin{align}
    \Psi_\mathcal{W}(\vec{\sigma}) = \text{exp}[\text{i}\phi_{\mathcal{W}}(\vec{\sigma})] \sqrt{p_{\mathcal{W}}(\vec\sigma)}\label{eq:ansatz}
\end{align}
The inclusion of the complex phases is the key difference between the wavefunctions used in this work and those used in Ref.~\cite{moss_leveraging_2025}. 

We employ RNN wavefunctions because of their recurrent nature. As depicted in \Cref{fig:RNN}, the same recurrent cell and dense layers, and thus the same set of weights $\mathcal{W}$ are used across the whole lattice. When weights are shared in this way, one forward pass of the RNN wavefunction involves repeatedly applying the same cell and dense layers, hence the name \emph{recurrent} neural network. Sharing the weights in this way allows us to define a variational wavefunction with a total number of parameters that does not explicitly depend on the number of spins $N$ in the system. Instead, the number of weights in the network is solely determined by the choice for the size of the hidden vectors $d_h$. Thus, $d_h$ is the single hyperparameter that controls the expressiveness of the ansatz for all system sizes.

\subsection{Symmetries}
\label{methods:symmetries}
A triangular lattice with periodic boundary conditions is symmetric under the point group $\mathcal{G}=C_{6v}$ which contains $|\mathcal{G}|=12$ symmetry transformations: six rotations over a $60^\circ$ degree angle and a single reflection. For a $L\times L$ triangular lattice with open boundaries, this symmetry is reduced to $\mathcal{G}=C_{2v}$. $C_{2v}$ which contains only a $180^\circ$ degree angle and a single reflection giving $|\mathcal{G}|=4$. Similar to previous works~\cite{Reisert2007equiv,nomura_helping_2021,schmitt2020quantum, hibat-allah_recurrent_2020, hibat-allah_supplementing_2024}, we incorporate symmetries in the variational ansatz by symmetry averaging the magnitude of the amplitudes of the wavefunction over the point group $\mathcal{G}$,
\begin{align}
     p_\mathcal{W}^\prime(\vec{\sigma}) = \frac{1}{|\mathcal{G}|} \sum_{\mathcal{T} \in \mathcal{G}} ^{|\mathcal{G}|}p_\mathcal{W}(\mathcal{T}\vec{\sigma}).
     \label{eq:sym_avg}
\end{align}
where $\mathcal{T}$ is a representation of a group element on the space of spin configurations $\vec{\sigma} \in\{\uparrow, \downarrow\}^N$.
In addition to symmetry averaging the magnitude of the amplitudes, one can also average the phases of the wavefunction over the group $\mathcal{G}$~\cite{hibat-allah_recurrent_2020}. However, it was observed in Ref.~\cite{Reh_optimizing_2023} that this method of phase averaging can fail to correctly capture the correct phases of states with a non-trivial sign structure.
In our case, we find that phase averaging does not improve our results and is numerically unstable for periodic systems. We investigate the reasons for these instabilities in detail in~\Cref{app:phase_avg}.

For the Heisenberg antiferromagnet defined on a bipartite lattice, the ground state is a singlet state with total spin equal to zero~\cite{marshall_antiferromagnetism_1955,lieb_ordering_1962}. A singlet state with zero total spin also has zero magnetization in the $z$-basis. This result concerning the nature of the finite-size ground state of the Heisenberg antiferromagnet does not extend to non-bipartite lattices; however, it has been heuristically observed that the finite-size ground state of the TLAHM is also a singlet state with zero total spin and zero magnetization in the $z$-basis~\cite{bernu_signature_1992,bernu_exact_1994}. As such, we enforce U(1) symmetry by restricting our RNN wavefunctions to states with zero magnetization in the $z$-basis~\cite{hibat-allah_supplementing_2024,hibat-allah_recurrent_2020,morawetz_u1-symmetric_2021}. The incorporation of this symmetry was shown to improve the variational energies achieved with RNN wavefunctions in past studies of the TLAHM~\cite{hibat-allah_supplementing_2024}.

\subsection{Iterative retraining}
\label{methods:retraining}

We define our RNN wavefunctions with shared weights such that the number of variational parameters is independent of the system size. This property allows us to use the optimized weights from a ground-state simulation for one system size as the initial weights for a ground-state simulation for a larger system size, a training procedure known as ``iterative retraining''~\cite{roth_iterative_2020,hibat-allah_supplementing_2024,moss_leveraging_2025}. This transfer-learning approach allows us to apply what is learned in one simulation (for example the correlations between spins) to the next simulation, which reduces the amount of information that must be learned from scratch for the simulation of the larger system. As a result, we are able to reduce the number of training steps as we grow the lattice size~\cite{roth_iterative_2020}. To control the number of training steps, we use a training schedule that is given by the parameterized function:
\begin{align}
    N_\text{steps}(L,s,r;L_0,C,F) = s\times\left[ C\text{exp}(-r(L-L_0)) + F\right],
    \label{eq:schedule}
\end{align}
where $s$ is an overall \emph{scale} factor, $r$ is the \emph{rate} at which the number of steps per $L$ decays. We fix $C$, which adjusts the number of training steps for the smallest system size $L_0$, and $F$, which determines the number of training steps for very large lattice sizes. The smallest system size we consider is $L_0=6$. The values of $C$ and $F$ are discussed in \Cref{app:training_details}. We note that when $s = 1.0$ and $r = 0.475$, the training schedule given by \Cref{eq:schedule} closely matches the schedule employed in previous iterative retraining studies~\cite{hibat-allah_supplementing_2024}.

\begin{figure}
    \centering
    \includegraphics[width=\linewidth]{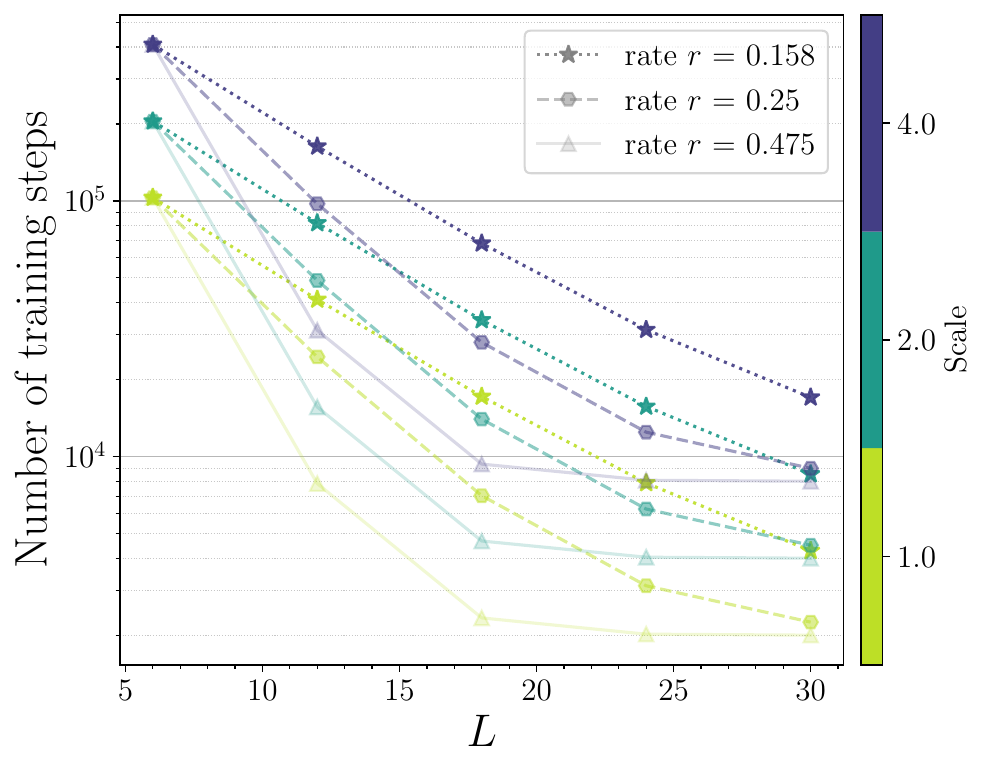}
    \caption{The number of training steps used in the optimization for each system size, as determined by our parameterized training schedule defined in \Cref{eq:schedule}. We consider three different scales and three rates. The colors and markers are used to indicate these two parameters respectively. 
    }
    \label{fig:schedules}
\end{figure}

\Cref{fig:schedules} displays the main schedules we study in this work. 
More details about how these training schedules impact the overall runtime of our simulations can be found in \Cref{app:runtimes}.

\section{Results and discussion}
\label{sec:results}

The methods described above form a powerful toolbox that allows us to simulate the ground states of the TLAHM defined on very large lattices. Using the results from those simulations, we are able to estimate ground-state properties of the TLAHM in the thermodynamic limit through a finite-size scaling study.
In particular, we examine the variational energies and estimates of the sublattice magnetization obtained from our trained ans\"{a}tze for system sizes up to $L=30$.

We focus on our results for the TLAHM with periodic boundary conditions, which has been studied more thoroughly with analytical approaches and other numerical methods. We performed a similar analysis for the TLAHM on lattices with open boundary conditions, which both supports and expands on previous RNN results~\cite{hibat-allah_supplementing_2024}. Those results can be found in \Cref{app:obc}.

\subsection{Learning accurate ground states for $L=6$ with variational neural annealing}
\label{results:annealing}

The quality of the results obtained from iteratively retraining an RNN wavefunction is sensitive to the accuracy of the ground state learned for the smallest system size considered. Therefore, we investigate which methods introduced in \Cref{sec:methods} yield the most accurate ground-state energies for $L=6$, which is the system size that serves as the starting point for all of our subsequent iterative retraining.

\begin{figure}
    \centering
    \includegraphics[width=\linewidth]{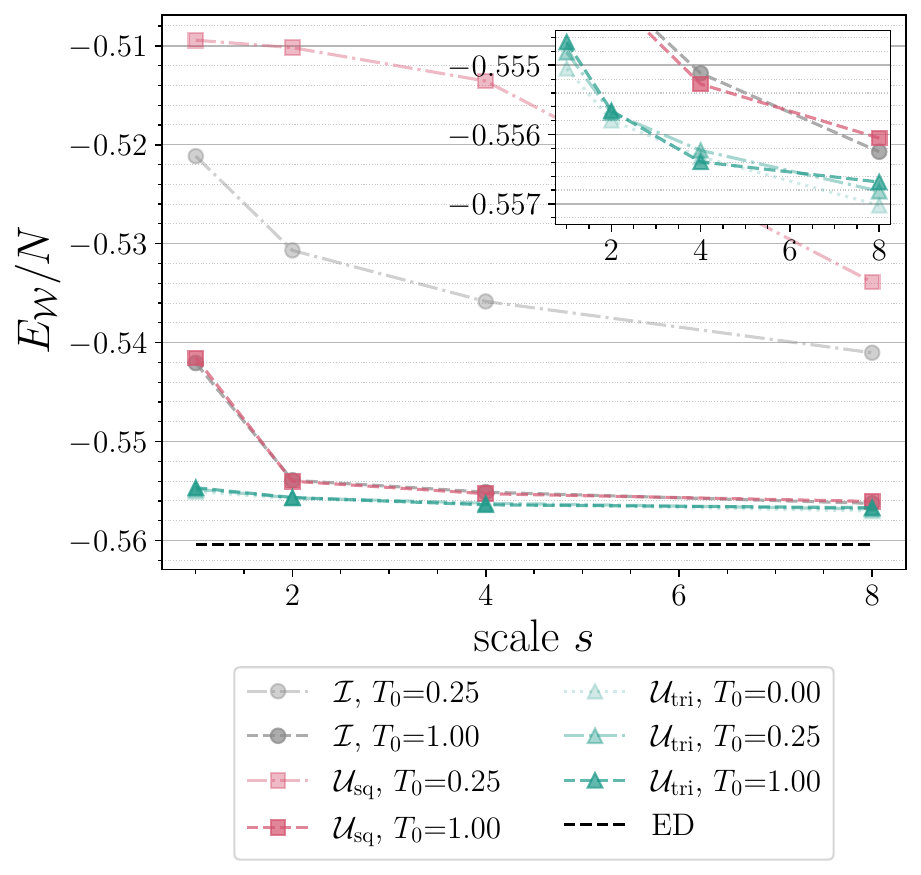}
    \caption{The energy per spin of the ground state of the TLAHM for $L=6$. We compare results from simulations performed with different local unitary basis transformations applied to the Hamiltonian: no basis transformation, $\mathcal{I}$, the basis transformation given by the MPSR, $\mathcal{U}_\text{sq}$ defined in \Cref{eq:sq_sign_rule}, and the 120$^{\circ}$ transformation, $\mathcal{U}_\text{tri}$ defined in \Cref{eq:tri_sign_rule}. Furthermore, we examine how the initial annealing temperature $T_0$ and the scale $s$ from \Cref{eq:schedule}, which combine to realize different annealing schedules, impacts the accuracy of our simulations. The inset shows a zoomed-in view of the results obtained when $\mathcal{U}_\text{tri}$ is employed. The ground-state energy obtained with exact diagonalization (ED) is shown for reference~\cite{bernu_exact_1994}.}
    \label{fig:L=6}
\end{figure}

\Cref{fig:L=6} shows how the annealing schedule, which is determined by the initial annealing temperature $T_0$ and the scale $s$ from \Cref{eq:schedule}, impact the accuracy of our simulations when different local unitary transformations are applied to the TLAHM Hamiltonian defined by \Cref{eq:hamiltonian}. 
As a baseline, we consider the case when no local unitary transformation, or equivalently the identity $\mathcal{I}$, is applied to the original Hamiltonian. We test the local unitary given by the MPSR, $\mathcal{U}_\text{sq}$,
which was employed in past attempts to study the TLAHM with iteratively retrained RNN wavefunctions~\cite{hibat-allah_supplementing_2024}. We also study the effects of the $120^\circ$ transformation, $\mathcal{U}_\text{tri}$, which is the local unitary transformation that is most tailored to the TLAHM ground state.

If we apply the identity $\mathcal{I}$ (no transformation) or $\mathcal{U}_\text{sq}$ (given by the MPSR) to the original TLAHM Hamiltonian, the VNA schedule has a significant impact on the quality of the simulation results.
For low initial annealing temperatures, when either of these two transformations is employed, our RNN wavefunctions are unable to reach accurate variational energies.
Interestingly, for $T_0=0.25$, the variational energies are worse when we use $\mathcal{U}_\text{sq}$ than when no transformation is used.
In both cases, however, we observe that a larger initial annealing temperature yields significantly lower variational energies. In fact, for $T_0=1.0$, the improved variational energies are roughly the same for both transformations. 
It is possible that $T_0=0.25$ is not a large enough initial annealing temperature to realize an exploration phase during the VNA, but that $T_0=1.0$ is, and that the exploration phase is crucial for overcoming the bias associated with the basis in which the Hamiltonian is defined.  
We also observe that annealing from the initial temperature $T_0$ to $T=0$ over a larger number of annealing steps, which is the result of larger scales $s$, further improves the results.

Notably, when the $120^\circ$ transformation is employed, the results of our simulations appear to be independent of the annealing schedule, since the initial temperature does not impact the final accuracy of the variational energy. These results can be seen more clearly in the inset of \Cref{fig:L=6}. Indeed, if the initial temperature is $T_0=0$, meaning the RNN wavefunction is trained using standard variational optimization, the variational energies appear to be very close to those obtained when $T_0>0$ and VNA is used. While increasing the scale $s$ does appear to improve the variational energies of our simulations, we believe this is simply due to the longer training time rather than a slower annealing process, since the initial temperature appears to have a negligible effect. The same systematic improvement with increased scale $s$ was observed for the square-lattice antiferromagnetic Heisenberg model, where traditional variational optimization was used to iteratively retrain RNN wavefunctions~\cite{moss_leveraging_2025}.

To summarize, our results support the idea that the ability to learn an accurate representation of a ground-state wavefunction is basis-dependent~\cite{park_geometry_2020,yang_when_2024}.
For the TLAHM, simulations when the $120^\circ$ transformation is applied to $\hat{H}$ are much more effective than when no transformation, $\mathcal{I}$, or a transformation that is tailored to a different ground state, $\mathcal{U}_{\text{sq}}$, are used. 
However, it is important to acknowledge that $\mathcal{U}_\text{tri}$ is inspired by knowledge of the ground state, which is not always available. 
Our results demonstrate that VNA is an effective optimization technique, which, given the right initial temperature, can significantly improve the ability to learn an accurate ground-state wavefunction even if the Hamiltonian is defined in a suboptimal basis, i.e., if the computational basis is not optimal or if a suboptimal basis transformation is applied to the Hamiltonian. Therefore, VNA could then be an effective approach in the case where the optimal basis for a given Hamiltonian is unknown.

\subsection{Finite-size scaling}

Starting with the ground states obtained for the TLAHM with $L=6$, we iteratively retrain our RNN wavefunctions for lattices up to $L=30$. Based on the results presented in the previous section, we employ the $120^\circ$ transformation, $\mathcal{U}_\text{tri}$, and take $T_0=1$ as our initial annealing temperature. 
For each system size, we can estimate the ground-state energy and the squared sublattice magnetization from the trained RNN wavefunction. The finite-size estimates of these observables can then be extrapolated to the thermodynamic limit. 
We benchmark our results against the values found in the literature for the ground-state energy and the squared sublattice magnetization in the thermodynamic limit. The first set of reference values was obtained by performing a finite-size scaling of large-scale DMRG simulations performed on matrix product states defined on cylinders~\cite{huang_magnetization_2024}. The second set of reference values was obtained with variationally-optimized infinite projected entangled-pair states (iPEPS)~\cite{hasik_incommensurate_2024}, a tensor network ansatz that directly parametrizes the ground state in the thermodynamic limit.

\subsubsection{Energy}
\label{results:energy}

First, we examine the variational energies of our iteratively retrained RNN wavefunctions. If the error of the variational energy is smaller than the energy gap between the ground state and the first excited state, then one can conclude that the largest contribution to the variational wavefunction is the true ground-state wavefunction~\cite{becca_quantum_2017}. 
In order to decide this, however, one must have access to reference energies in order to compute the energy error, in addition to estimates of the energy gap. Exact energies and estimates of the gap are not accessible for large system sizes, and for the TLAHM, we do not have estimates of these quantities that are considered numerically exact. The absence of these estimates reflects the difficulty of simulating the ground state of the TLAHM. 

Without reference values for finite-size energies, the best way to assess the quality of our results is to perform a finite-size scaling using our estimates of the ground-state energy for each system size. We can then compare our extrapolated estimates of the ground-state energy in the thermodynamic limit with values from the literature. For systems with periodic boundary conditions, we use the following scaling form for the energy~\cite{bernu_signature_1992,bernu_exact_1994,capriotti_long-range_1999} 
\begin{align}
    E(L) = E_\infty + \frac{e_1}{L^3} + \mathcal{O}\left(\frac{1}{L^4}\right).
    \label{eq:energy_scaling}
\end{align}

\begin{figure}
    \centering
    \includegraphics[width=\linewidth]{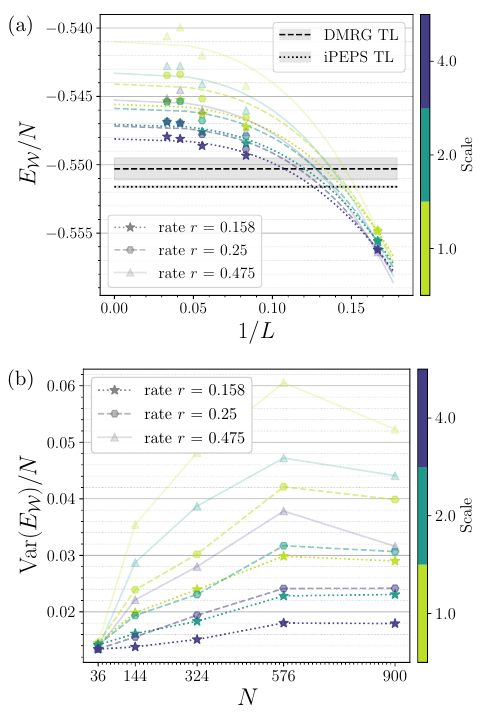}
    \caption{(a) The variational energies for all system sizes obtained from the RNN wavefunctions optimized according to each of the iterative retraining schedules shown in \Cref{fig:schedules}. These energies are plotted according to $1/L$ for easier viewing, but they are fit according to the scaling form defined by \Cref{eq:energy_scaling}. The reference values of the ground-state energy in the thermodynamic limit (TL) are shown for comparison. The dashed line is the value from DMRG simulations using MPS with cylindrical boundary conditions~\cite{huang_magnetization_2024}. The dotted line is the value from variationally-optimized iPEPS~\cite{hasik_incommensurate_2024}. Each of our variational energies is estimated with $10\times10^3$ samples.
    (b) The variances of the final variational energies shown in (a) plotted as a function of system size $L$.}
    \label{fig:energies_peri}
\end{figure}

\Cref{fig:energies_peri}(a) shows the variational energies obtained for all system sizes from the RNN wavefunctions trained according to the schedules displayed in \Cref{fig:schedules}. We also show the extrapolation of these finite-size energies according to \Cref{eq:energy_scaling}. Our results show systematic improvement, i.e., lower energies, with larger scales $s$ or slower rates $r$, both of which increase the overall training time. Not only do the finite-size variational energies improve, but the estimates of the ground-state energy in the thermodynamic limit approach the reference values for longer training schedules.

The variance of the variational energy can be interpreted as a measure of how close a variational wavefunction is to an energy eigenstate~\cite{wu_variational_2024,Gross_criterion_1990,Assaraf_zerovar_2003}, making it another useful quantity for determining the quality of our results. \Cref{fig:energies_peri}(b) shows the variances of the variational energies shown in (a). 
We observe that in addition to lowering the energies, longer training times systematically reduce the variance of the variational energy.
Another metric, known as the V-score~\cite{wu_variational_2024}, combines the variational energy and the associated variance as a way to assess the quality of variational results in a problem-independent way. We examine the V-scores of our simulation results in \Cref{app:vscores}.

Ultimately, our results in \Cref{fig:energies_peri}(a) and (b) illustrate how our RNN wavefunctions trained according to the different schedules constitute a converging sequence of states. It is common to take such a sequence of states and extrapolate the corresponding variational energies to the zero-variance limit~\cite{nomura_restricted_2017,kwon_effects_1998,becca2000stabilitydwavesuperconductivitytj,Sorella2001,hu_direct_2013,fu_variance_2023}. 
For a given system size $L$, an estimate of the zero-variance energy can be extracted with a linear fit through the variational energies and the corresponding variances obtained from each RNN wavefunction. See \Cref{app:zero-var} for an example of our zero-variance extrapolation for $L=30$. After performing this linear fit through our best variational results~\footnote{We perform our zero-variance extrapolation using the results from simulations with $s=1.0,2.0,4.0,r=0.158$ in addition to results from simulations with $s=2.0,4.0,r=0.25$. These simulations produced the best variational energies and the lowest variances of those variational energies, as seen in \Cref{fig:energies_peri}.}, we improve the fit using the wild bootstrap method~\cite{wu_bootstrap}. The wild bootstrap method is a resampling technique used to improve the statistical accuracy of a regression. 

\begin{figure}
    \centering
    \includegraphics[width=\linewidth]{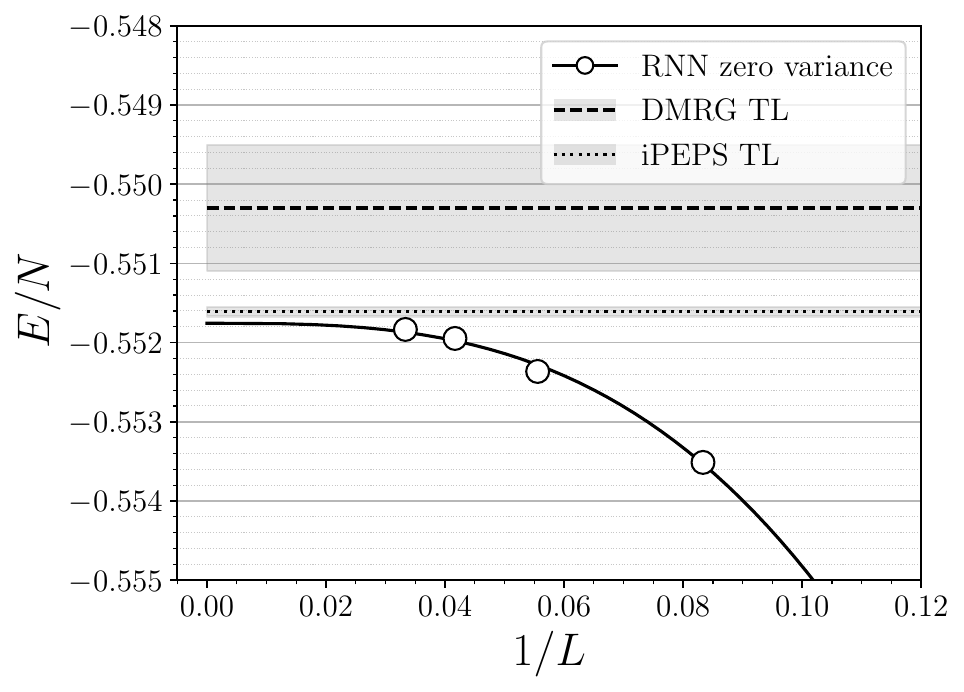}
    \caption{Improved estimates of the ground-state energies for finite sizes obtained from extrapolating the energies shown in \Cref{fig:energies_peri}(a) to their zero-variance limit for each system size $L$. These zero-variance energies, excluding the value for $L=6$, are fit according to the scaling form defined by \Cref{eq:energy_scaling}. We plot these values as a function of $1/L$ for easier viewing. The reference values of the ground-state energy in the thermodynamic limit (TL) are shown for comparison. The dashed line is the value from DMRG simulations using MPS with cylindrical boundary conditions~\cite{huang_magnetization_2024}. The dotted line is the value from variationally-optimized iPEPS~\cite{hasik_incommensurate_2024}. }
    \label{fig:zero_var_energy}
\end{figure}

The bootstrapped zero-variance energies obtained from our results and $10^3$ bootstraps (instances of resampling a data point according to its mean value and variance) are shown in \Cref{fig:zero_var_energy}.
These improved, zero-variance estimates of the finite-size energies can then be used to perform an improved finite-size scaling. 
We note that we did not include the zero-variance energy for $L=6$ in our finite-size scaling because our variational energies and their corresponding variances are all very close together, making it difficult to reliably fit a line through the data. From our zero-variance energies, we estimate the ground-state energy to be $E_\infty = -0.5517569(9)$, which is lower than the reference DMRG value $E_\infty^{\text{DMRG}} = -0.5503(8)$~\cite{huang_magnetization_2024} by  $1\times10^{-3}$ and within $1\times10^{-4}$ of the reference iPEPS value $E_\infty^{\text{iPEPS}} = -0.55161(6)$~\cite{hasik_incommensurate_2024}, which is a strict upper-bound on the true ground-state energy in the thermodynamic limit. 
Importantly, our bootstrapped zero-variance energies are no longer variational, meaning it is possible that they are below the true ground-state energy for a given system size. As a result, the extrapolated value of the ground-state energy in the thermodynamic limit could also be below the true ground-state energy. However, the close agreement between our extrapolated energy and the value obtained with iPEPS suggests that any deviation below the true ground-state energy is small (at most $1\times10^{-4}$).

Although we focus only on periodic boundaries in this section, we note that we were also able to extract an accurate estimate of the ground-state energy in the thermodynamic limit from our simulations for lattices with open boundary conditions.
Interestingly, we obtained this accurate estimate of the ground-state energy in the thermodynamic limit from the finite-size scaling of results from a \emph{single} simulation that did not require much training time (scale $s = 1.0$, rate $r=0.158$). See \Cref{app:obc} for more details.

\subsubsection{Sublattice magnetization}
\label{results:magnetization}

In addition to the variational energies, we examine the correlations captured by our optimized RNN wavefunctions. The correlations provide important information about the nature of a quantum state. For instance, the order parameter that defines a particular phase of matter is typically some function of the correlations. For the TLAHM, correlations can be used to compute the order parameter that defines the long-range antiferromagnetic order of the ground state, the sublattice magnetization.
The first step is to measure the real-space correlations,
\begin{align}
    C(i,j) = \left\langle \vec{S}_i\cdot\vec{S}_j \right\rangle,
    \label{eq:real_space_corr}\\
    C^z(i,j) = 3\left\langle S^{z}_i S^{z}_j \right\rangle.
    \label{eq:real_space_corr_z}
\end{align}
The Heisenberg model and its finite-size ground states are SU(2) symmetric, which means that quantities computed with the correlations defined by \Cref{eq:real_space_corr} should be equal to the same quantities computed with the correlations defined by \Cref{eq:real_space_corr_z}. This fact is often exploited, and observables are computed only with $C^z(i,j)$ correlations, which are diagonal in the computational basis and thus typically cheaper to compute using Monte Carlo techniques. However, this substitution is only valid if the optimized variational state has indeed found a state that is SU(2) symmetric.

\begin{figure}
    \centering
    \includegraphics[width=\linewidth]{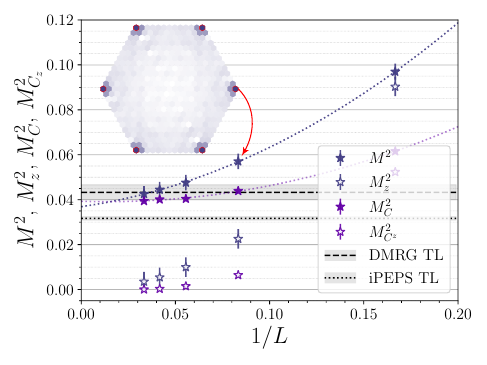}
    \caption{Our estimates of the squared sublattice magnetization scaled as a function of $1/L$. We show estimates of $M^2$ and $M^2_C$ to confirm that our estimates extrapolate to the same value in the thermodynamic limit. The closed markers correspond to the values of $M^2$ and $M^2_C$ estimated using the correlations defined in \Cref{eq:real_space_corr} and open markers correspond to the values of $M^2_z$ and $M^2_{C^z}$, which are estimated using the correlations defined by \Cref{eq:real_space_corr_z}. 
    The reference values for the squared sublattice magnetization in the thermodynamic limit (TL) are shown for comparison. The dashed line is the value from DMRG simulations using MPS with cylindrical boundary conditions~\cite{huang_magnetization_2024}. The dotted line is the value from variationally-optimized iPEPS~\cite{hasik_incommensurate_2024}. 
    The inset shows all of the  momentum-space correlations captured by our RNN wavefunction for $L=12$. The points in momentum space corresponding to the expected ordering wavevectors are highlighted in red.}
    \label{fig:M_scaling_peri}
\end{figure}

The real-space correlations can be used to compute the momentum-space correlations
\begin{align}
    S^{(z)}(L,\vec{q}) = \frac{1}{L^2} \sum_{i,j} e^{i\vec{q}\cdot\vec{r}} C^{(z)}(i,j),
    \label{eq:structure_factor}
\end{align}
where $\vec{r} = \vert\vec{r}_i-\vec{r}_j\vert$. 
The momentum-space correlations can be computed with the correlations defined in either \Cref{eq:real_space_corr} or \Cref{eq:real_space_corr_z}, as indicated by the superscript $(z)$. For ordered ground states, the momentum-space correlations will show peaks at the points in momentum space that correspond to the ordering wavevectors. In the case of the TLAHM ground state, which has $120^\circ$ magnetic order, those peaks appear at $\vec{q}=(\frac{4\pi}{3},0)$ and the six possible $60^\circ$ rotations of that vector. The inset of \Cref{fig:M_scaling_peri} shows the momentum-space correlations captured by a trained RNN wavefunction for $L=12$. The peak values are exactly at the points corresponding to the expected ordering wavevectors, which are outlined in red. 

The order parameter for the antiferromagnetic long-range order is the sublattice magnetization, which should remain finite in the thermodynamic limit. Typically, the squared sublattice magnetization $M^2$ is used for finite-size scaling. For periodic lattices, $M^2$ can be estimated in two ways. The first approach involves scaling the peak value of the momentum-space correlations, \Cref{eq:structure_factor} evaluated at $\vec{q} = (\frac{4\pi}{3},0)$, by the size of the system,
\begin{align}
    M^2_{(z)}(L) = \frac{S^{(z)}\big(L,\vec{q} = (\frac{4\pi}{3},0)\big)}{L^2}.
    \label{eq:M}
\end{align}
The same quantity can be estimated using the real space correlations between spin pairs that are separated by the longest separation vector,
\begin{align}
    M^2_{C^{(z)}}(L) = C^{(z)}(L/2,L/2).
    \label{eq:M_C}
\end{align}
Finite-size estimates of $M^2$ and $M^2_C$ should extrapolate to the same value of the squared sublattice magnetization in the thermodynamic limit.

\Cref{fig:M_scaling_peri} shows the values for the squared sublattice magnetization estimated from our longest and most accurate simulations (scale $s=4.0$, rate $r=0.158$). We show the values for both \Cref{eq:M} and \Cref{eq:M_C}.
Each real-space correlation was estimated with $10\times10^3$ samples. In order to estimate the squared sublattice magnetization in the thermodynamic limit, we fit a second-order polynomial in $1/L$ to our finite-size estimates of $M^2$ and $M^2_C$. Notably, the second-order fit has three parameters, meaning it was necessary to obtain all of the finite-size estimates for $M^2$ in order to obtain a reliable fit through the data. The calculation of $M^2$ scales as $\mathcal{O}(N^3)$ since \Cref{eq:structure_factor} involves a sum over all two-point correlations and performing inference with our RNN wavefunction scales as $\mathcal{O}(N)$. After performing the initial second-order fit, we use the wild bootstrap method~\cite{wu_bootstrap} with $10\times10^3$ bootstraps to improve the statistical accuracy of our extrapolated estimates of the squared sublattice magnetization. From this finite-size scaling of our estimates of $M^2$ we get an estimate of the sublattice magnetization in the thermodynamic limit of $M_\infty = 0.192(2)$ and from extrapolating our estimates of $M^2_C$ we get $M_\infty = 0.198(2)$.
These estimates are in good agreement with one another, and likely with more finite-size estimates for larger system sizes, the extrapolated values would converge. The estimate for the sublattice magnetization in the thermodynamic limit obtained from extrapolating $M^2_C$ is within error bars of the reference DMRG value $M_\infty^\text{DMRG} = 0.208(8)$~\cite{huang_magnetization_2024}. The reference iPEPS value $M_\infty^\mathrm{iPEPS}$ is considerably lower than our estimates~\cite{hasik_incommensurate_2024}, but a direct comparison is difficult because $M_\infty^{\mathrm{iPEPS}}$ is not estimated in the same way as $M_\infty^{\mathrm{DMRG}}$ and our estimates of $M_\infty$.
A variational estimate of any observable, aside from the energy, is not considered an upper-bound to the true ground-state value of that observable. This freedom is manifest in the outstanding discrepancy between the estimated values of the sublattice magnetization coming from different numerical techniques, including our own. 

An important observation about the results in \Cref{fig:M_scaling_peri} is the disagreement between $M^2$ and $M^2_z$ (and similarly between $M^2_C$ and $M^2_{C^z}$). The disagreement between the two quantities reveals that our RNN wavefunction has not learned an SU(2)-symmetric state. We believe that our variational wavefunction has learned a superposition of the states in Anderson's tower of states $\vert 0 \rangle$, which is indeed a symmetry-broken state.
All of the eigenstates in $\vert 0 \rangle$ have different values of the total spin, with the true finite-size ground state being a singlet with zero total spin. Therefore, a superposition of the states in $\vert0\rangle$ would have a non-zero value of the total spin $\langle\vec{S}^2\rangle$.
However, the states in $\vert0\rangle$ notably exhibit the same value of the squared sublattice magnetization~\cite{bernu_signature_1992,bernu_exact_1994}. Therefore, for a given system size, a superposition of the states in $\vert 0 \rangle$ should yield an estimate of the squared sublattice magnetization that is close to the estimate obtained from the true finite-size ground state. An extrapolation to the thermodynamic limit using estimates from such superpositions should then produce an accurate estimate of the squared sublattice magnetization in the thermodynamic limit. 

Given that (i) our extrapolated values for the squared sublattice magnetization are in good agreement with the values from the literature, (ii) our variational wavefunctions are not SU(2)-symmetric, and (iii) the states learned by our RNN wavefunctions have $\langle\vec{S}^2\rangle>0$ (see \Cref{app:total_spin}), we conclude that the state learned by our RNN wavefunctions is dominated by the states in Anderson's tower of states $\vert 0 \rangle$. 
Other quantities, such as the entanglement entropy~\cite{kulchytskyy_detecting_2015}, might allow us to detect whether the states learned by our RNN wavefunctions contain contributions from higher excited states that are not in Anderson's tower of states.

\section{Conclusion}
\label{sec:conclusion}

This study benchmarks the performance of iteratively retrained two-dimensional RNN wavefunctions on the TLAHM. Because of the ability to iteratively retrain our ans\"{a}tze, we are able to simulate ground states for very large two-dimensional systems with lattice sizes up to $L=30$. From these results, we are able to extract estimates of ground-state properties in the thermodynamic limit through finite-size scaling. Our thermodynamic limit estimates of the ground-state energy and the squared sublattice magnetization are in close agreement with values from the literature. In \Cref{app:comparison} we directly compare our estimated quantities with available reference values.

Our results show systematic improvement with increased training time, which is controlled by a parameterized training schedule. The ability to effectively use this schedule, which hinges on the success of iterative retraining, helps moderate the overall runtime of our simulations. Even though our longest simulation required a manageable amount of compute, we required more training steps for larger lattice sizes, i.e., rate $r=0.158$, resulting in longer runtimes than the simulations performed for the square-lattice antiferromagnetic Heisenberg model in Ref.~\cite{moss_leveraging_2025} (where the smallest rate considered was $r=0.25$). Even with the increased resource usage, our results for the ground-state properties of the TLAHM are not as accurate as those obtained for the square-lattice, indicating that learning the ground states of the TLAHM is a more challenging optimization task that requires even more compute. 
This increased difficulty is manifest in the V-scores of our variational results for the TLAHM, which are much larger (worse) than those obtained with the same methods for the square-lattice antiferromagnetic Heisenberg model~\cite{moss_leveraging_2025} (see \Cref{app:vscores}). 

The most appreciable difference between the TLAHM and its square-lattice counterpart is the inability to make the Hamiltonian stoquastic with a local unitary transformation. VMC does not provably suffer from a negative sign problem like QMC methods, but it is possible that the increased difficulty in simulating the TLAHM that we observe is a consequence of the non-definite sign of the wavefunction amplitudes, what we refer to as a non-trivial sign structure. 
For instance, in \Cref{fig:L=6} we show that the ability to learn the ground state of the TLAHM is basis-dependent. The sign structure of a wavefunction is also basis-dependent, so it is reasonable to hypothesize that the ability to learn the ground state of the TLAHM is tied to the sign structure of the ground-state wavefunction in the chosen basis.
Furthermore, it is possible that the non-trivial sign structure learned by the RNN wavefunction does not generalize from one system size to the next, impacting the quality of our results when we iteratively retrain our RNN wavefunctions. A more systematic investigation of the sign structures learned by our RNN wavefunctions would elucidate many of our results.

Despite recent results on the limited expressiveness of RNNs~\cite{bortone_impact_2024,yang_when_2024,jreissaty_2025}, our results indicate that RNN wavefunctions can be systematically improved with more training time, which suggests that expressiveness is not a bottleneck, but rather the optimization is.
We emphasize that all of the results presented here were obtained from RNN wavefunctions optimized with Adam: a first-order optimization method that is common in classical machine learning~\cite{kingma2017adammethodstochasticoptimization}. Stochastic Reconfiguration (SR) is a different optimization technique that is closely related to the imaginary time evolution of a quantum state. This method involves adjusting the gradients used during optimization according to the geometry of the optimization landscape. Many of the strongest results in the field of NQS have been achieved with SR~\cite{becca_quantum_2017} or its variants~\cite{chen_empowering_2024,rende_simple_2024}, but SR has not been effective when applied to RNN wavefunctions~\cite{donatella_autoregressive,lange2024neural}.  Indeed, it is an open question whether this optimization method can be used to successfully train RNN wavefunctions, and whether this will improve existing results. 
In some cases, ans\"{a}tze optimized with Adam reach equally accurate results as ans\"{a}tze optimized with SR in a comparable amount of training \emph{time}~\cite{malyshev_neural_2024}, since SR is more computationally demanding even in its modified forms~\cite{chen_empowering_2024,rende_simple_2024}. 
To this point, many of the aforementioned results obtained using SR or its variants required huge amounts of computation to obtain a single variational energy. For example, in Ref.~\cite{rende_simple_2024}, one simulation of the $J_1-J_2$ Heisenberg model on a $10\times 10$ square lattice took 4 days on twenty A100 GPUs, which is roughly 1,900 GPU hours. 
The longest simulation reported in this work took 1,700 GPU hours and produced energies for six different system sizes up to $30\times30$, albeit with more modern hardware (see \Cref{app:runtimes}).
So although SR might be a highly effective method for optimizing some NQS, it might limit one to architectures that are not able to be iteratively retrained and are therefore more difficult to scale.

In conclusion, we have demonstrated that RNN wavefunctions are naturally scalable and able to successfully generalize, allowing us to accurately simulate ground states of the TLAHM for very large lattices without prohibitive demands for computational resources. Our finite-size scaling of the ground-state properties obtained from our simulations show good agreement with values from the literature, confirming that our methods are effective even for frustrated Hamiltonians that host ground states with non-trivial sign structures.

\section{Acknowledgments}
\label{sec:acknowledgments}
SM would like to acknowledge Chris Roth and Estelle Inack for helpful discussions throughout the completion of this work. SM would also like to thank Juraj Hasik for pointing out Ref.~\cite{hasik_incommensurate_2024}.

We acknowledge financial support from the
Natural Sciences and Engineering Research Council of
Canada (NSERC) and the
Perimeter Institute. Research at Perimeter Institute is
supported in part by the Government of Canada through
the Department of Innovation, Science and Economic
Development Canada and by the Province of Ontario
through the Ministry of Economic Development, Job
Creation and Trade. RW acknowledges support from the Flatiron Institute. The Flatiron Institute is a division of the Simons Foundation. 

This work involved many large-scale simulations.
SM and RW thank the Perimeter Institute and Flatiron Institute Scientific Computing Center for computational resources and technical support. Resources used in preparing this research were provided, in part, by the Province of Ontario, the Government of Canada through CIFAR, and companies sponsoring the Vector Institute \url{https://vectorinstitute.ai/#partners}. Additional resources were provided by the Shared Hierarchical Academic Research Computing Network (SHARCNET) and
the Digital Research Alliance of Canada. 

\section{Code Availability}
\label{sec:code}
Our implementation of the presented methods, all scripts needed to reproduce our results, and most of the data needed to reproduce the figures in this manuscript are openly available on GitHub \url{github.com/mschuylermoss/HeisenbergRNN}~\cite{Moss2025git}. Some data files are too large to host on GitHub, but can be shared upon request. Our code relies on TensorFlow~\cite{tensorflow2015-whitepaper}, NumPy~\cite{harris2020array}, and Matplotlib~\cite{Hunter:2007}. 

\clearpage
\appendix

\section{Details of square and triangular sign rules}
\label{app:sign_rules}

\begin{figure}
    \centering
    \includegraphics[width=0.9\linewidth]{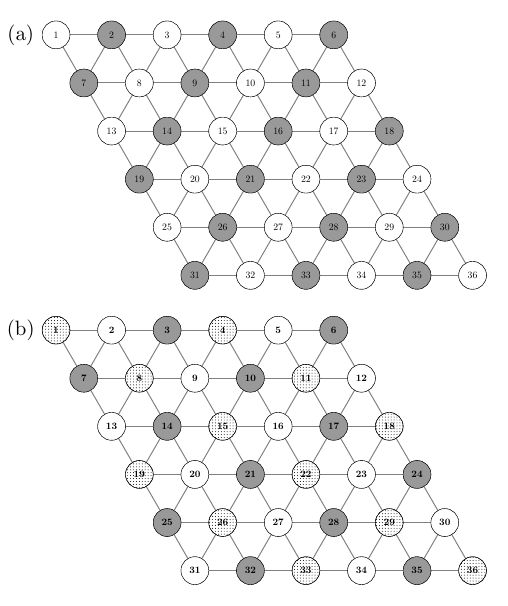}
    \caption{ A covering of a $6\times6$ triangular lattice with (a) the two sublattices A$_\text{sq}$ and B$_\text{sq}$ and (b) the three sublattices A$_\text{tri}$, B$_\text{tri}$, and C$_\text{tri}$. The local unitary transformation given by the MPSR, $\mathcal{U}_\text{sq}$, consists of local rotations that act on the spins in B$_\text{sq}$. The $120^\circ$ transformation, $\mathcal{U}_\text{tri}$, consists of local rotations that act on the spins in B$_\text{tri}$ and C$_\text{tri}$.}
    \label{fig:sublattices}
\end{figure}

In this work we employ two local unitary transformations: $\mathcal{U}_\text{sq}$ defined by \Cref{eq:sq_sign_rule}, which is given by the Marshall-Peierls sign rule (MPSR), and $\mathcal{U}_\text{tri}$ defined by \Cref{eq:tri_sign_rule}, which we refer to as the $120^\circ$ transformation. These unitary transformations are composed of single-spin rotations, where the rotation applied to a given spin depends on the sublattice to which that spin belongs. For $\mathcal{U}_\text{sq}$, we define two sublattices and for $\mathcal{U}_\text{tri}$, we define three. These sublattices are depicted in \Cref{fig:sublattices}. 

Partitioning the system into sublattices A$_\text{sq}$ and B$_\text{sq}$ shown in \Cref{fig:sublattices}(a), $\mathcal{U}_\text{sq}$ rotates all spins on sublattice B$_\text{sq}$ by $\pi$ around the $z$-axis. This local unitary transformation famously makes the square-lattice antiferromagnetic Heisenberg model stoquastic~\cite{marshall_antiferromagnetism_1955,capriotti_quantum_2001}. In order to see the effect of this transformation on the TLAHM defined by \Cref{eq:hamiltonian}, we separate the off-diagonal terms of the Hamiltonian based on the sublattice that they act on,
\begin{align*}
    \hat{H} =  \sum_{\langle ij \rangle} \hat{S}^{z}_i\hat{S}^{z}_j  &+ \frac{1}{2}\sum_{i,j \in \text{A}_{\text{sq}}}(\hat{S}^{+}_i\hat{S}^{-}_j + \hat{S}^{-}_i\hat{S}^{+}_j) \\
    &+ \frac{1}{2}\sum_{i,j \in \text{B}_{\text{sq}}} (\hat{S}^{+}_i\hat{S}^{-}_j + \hat{S}^{-}_i\hat{S}^{+}_j)\\
    & + \frac{1}{2}\sum_{\substack{i\in\text{A}_{\text{sq}},\\j\in\text{B}_{\text{sq}}}} (\hat{S}^{+}_i\hat{S}^{-}_j + \hat{S}^{-}_i\hat{S}^{+}_j).
\end{align*}
There are three different types of interactions: interactions between spins that are both in A$_\text{sq}$, interactions between spins that are both in B$_\text{sq}$, and interactions between spins that belong to different sublattices. For the first two types of interactions, $\mathcal{U}_\text{sq}$ has no effect. However, $\mathcal{U}_\text{sq}$ will contribute a negative sign to the term corresponding to interactions where the two spins belong to different sublattices. Therefore, the transformed Hamiltonian can be rewritten as
\begin{align*}
    \mathcal{U}_\text{sq}^\dagger\hat{H}\mathcal{U}_\text{sq} =  \sum_{\langle ij \rangle} \hat{S}^{z}_i\hat{S}^{z}_j  &+ \frac{1}{2}\sum_{i,j \in \text{A}_{\text{sq}}}(\hat{S}^{+}_i\hat{S}^{-}_j + \hat{S}^{-}_i\hat{S}^{+}_j) \\
    &+ \frac{1}{2}\sum_{i,j \in \text{B}_{\text{sq}}} (\hat{S}^{+}_i\hat{S}^{-}_j + \hat{S}^{-}_i\hat{S}^{+}_j)\\
    & - \frac{1}{2}\sum_{\substack{i\in\text{A}_{\text{sq}},\\j\in\text{B}_{\text{sq}}}} (\hat{S}^{+}_i\hat{S}^{-}_j + \hat{S}^{-}_i\hat{S}^{+}_j).
\end{align*}
As a result, two-thirds of the off-diagonal terms take on a leading minus sign. Notably, for the square-lattice antiferromagnetic Heisenberg model, all nearest-neighbor interactions are between spins on different sublattices, which is why this transformation makes all the off-diagonal terms in the Hamiltonian negative.

Alternatively, the $120^\circ$ transformation $\mathcal{U}_\text{tri}$ defined by \Cref{eq:tri_sign_rule} requires the spins on the lattice to be divided into three sublattices A$_\text{tri}$, B$_\text{tri}$, and C$_\text{tri}$. This unitary rotates all spins on sublattice B$_\text{tri}$ by $-\frac{2\pi}{3}$ around the $z$-axis and all spins on sublattice C$_\text{tri}$ by $+\frac{2\pi}{3}$ around the $z$-axis. We again separate the off-diagonal terms by the sublattice that they act on,
\begin{align*}
    \hat{H} = 
    \sum_{\langle ij \rangle} \hat{S}^{z}_i\hat{S}^{z}_z
    &+
     \frac{1}{2}\sum_{\substack{i\in\text{A}_{\text{tri}},\\j\in\text{B}_{\text{tri}}}}(\hat{S}^{+}_i\hat{S}^{-}_j + \hat{S}^{-}_i\hat{S}^{+}_j)\\
     &+
     \frac{1}{2}\sum_{\substack{i\in\text{B}_{\text{tri}},\\j\in\text{C}_{\text{tri}}}}
     (\hat{S}^{+}_i\hat{S}^{-}_j + \hat{S}^{-}_i\hat{S}^{+}_j)\\
     &+
     \frac{1}{2}\sum_{\substack{i\in\text{C}_{\text{tri}},\\j\in\text{A}_{\text{tri}}}}(\hat{S}^{+}_i\hat{S}^{-}_j + \hat{S}^{-}_i\hat{S}^{+}_j).
\end{align*}
In this case, each spin only interacts with spins that belong to a different sublattice, so there are three types of interactions. As a result, $\mathcal{U}_\text{tri}$ has the same effect on each of the terms corresponding to the three types of interactions. Under the transformation $\mathcal{U}_\text{tri}$ the original Hamiltonian then becomes,
\begin{align*}
    \mathcal{U}_\text{tri}^{\dagger}\hat{H}\mathcal{U}_\text{tri} = 
    \sum_{\langle ij \rangle} \hat{S}^{z}_i\hat{S}^{z}_z
    &-\frac{1}{4}\sum_{\langle ij \rangle} (\hat{S}^{+}_i\hat{S}^{-}_j + \hat{S}^{-}_i\hat{S}^{+}_j )\\
    &-\text{i}\frac{\sqrt{3}}{4} \sum_{\substack{i\in\text{A}_{\text{tri}},\\j\in\text{B}_{\text{tri}}}}(\hat{S}^{+}_i\hat{S}^{-}_j - \hat{S}^{-}_i\hat{S}^{+}_j)\\
     &-\text{i}\frac{\sqrt{3}}{4} \sum_{\substack{i\in\text{B}_{\text{tri}},\\j\in\text{C}_{\text{tri}}}}(\hat{S}^{+}_i\hat{S}^{-}_j - \hat{S}^{-}_i\hat{S}^{+}_j)\\
     &-\text{i}\frac{\sqrt{3}}{4} \sum_{\substack{i\in\text{C}_{\text{tri}},\\j\in\text{A}_{\text{tri}}}}(\hat{S}^{+}_i\hat{S}^{-}_j - \hat{S}^{-}_i\hat{S}^{+}_j),
\end{align*}
where we note that $-\text{i}\frac{\sqrt{3}}{4}(\hat{S}^{+}_i\hat{S}^{-}_j - \hat{S}^{-}_i\hat{S}^{+}_j) = -\frac{\sqrt{3}}{2}(\hat{S}^{x}_i\hat{S}^{y}_j - \hat{S}^{y}_i\hat{S}^{x}_j)$. From this new form of the Hamiltonian, it is not possible to say how many of the off-diagonal terms will have an overall negative sign. Heuristically, however, we see that this unitary transformation leads to lower energies and better results for iterative retraining. 

Enforcing U(1) symmetry requires that we have an even number of spins, and thus an even linear lattice length $L$, since we restrict our variational wavefunction to the $S_z=0$ sector.
This choice ensures that the two sublattices corresponding to $\mathcal{U}_\text{sq}$ are equally represented. When $\mathcal{U}_\text{tri}$ is employed, $L$ must also be divisible by 3, so that all three sublattices contain an equal number of spins. If each sublattice is not equally represented, then one sublattice will be energetically favored over the others. Therefore, we consider lattice lengths that are multiples of 6 when $\mathcal{U}_\mathrm{tri}$ is employed. We call these lattice sizes ``commensurate'' with the three-sublattice magnetic order present in the ground state of the TLAHM.

\section{Details of RNN wavefunctions}
\label{app:RNN}

RNNs process the information in the hidden vectors with the RNN cell. Gated recurrent units are the typical choice for RNN cell, as they help alleviate vanishing and exploding gradients~\cite{cho2014learning,chung2014empirical}. 
In a GRU cell three quantities are computed: a candidate hidden vector $\Tilde{\vec{h}}_{i,j}$, an update gate $\vec{u}_{i,j}$, and finally the output hidden vector $\vec{h}_{i,j}$. The boundary conditions impact how $\Tilde{\vec{h}}_{i,j}$ and $\vec{u}_{i,j}$ are calculated, since the RNN cells take more inputs when pseudo-periodic boundary conditions are employed.

For periodic boundary conditions, the cell takes four hidden vectors as inputs, so we employ a regular GRU cell. The input to the cell are the four hidden vectors $\vec{h}$ and one-hot representations of a single spin in the full configuration $\vec{\sigma}_i$ obtained from the (square-lattice) nearest-neighbor RNN cells. The hidden vectors and input spins are initialized with zeros and get assigned values as we progress through the RNN accordisng to the autoregressive sequence. Therefore, only hidden vectors and spins coming from nearest-neighbors that come earlier in the autoregressive sequence will contribute to the computations that occur in the cell, thus satisfying the requirements of  \Cref{eq:chain_rule_of_probabilities}.
The RNN cell computes a  candidate hidden vector and update gate as follows:
\begin{align*}
    \Tilde{\vec{h}}_{i,j} &= \text{tanh}\left(
    \left[\vec{h}_{\text{input}}; \vec{\sigma}_{\text{input}}]\right]
    \mathbf{W} + \vec{b}\right),
    \\
     \vec{u}_{i,j} &= \text{sigmoid}\left(
     \left[\vec{h}_{\text{input}}; \vec{\sigma}_{\text{input}}]\right]
    \mathbf{W}_g + \vec{b}_g\right).
\end{align*}
with
\begin{align*}
    \vec{h}_{\text{input}} &= [\vec{h}_{i-1,j};\vec{h}_{i,j-1};\vec{h}_{i+1,j};\vec{h}_{i,j+1}],\\
    \vec{\sigma}_{\text{input}} &= [\vec{\sigma}_{i-1,j};\vec{\sigma}_{i,j-1};\vec{\sigma}_{i+1,j};\vec{\sigma}_{i,j+1}],
\end{align*}
where $[\vec{a};\vec{b}]$ denotes the concatenation of two vectors. Note that for periodic boundary conditions, $\vec{h}_{i, L+1}\equiv \vec{h}_{i, 1}$ and $\vec{h}_{L+1, j}\equiv \vec{h}_{1, j}$ as well as $\vec{\sigma}_{i, L+1}\equiv \vec{\sigma}_{i, 1}$ and $\vec{\sigma}_{L+1, j}\equiv \vec{\sigma}_{1, j}$. 

When simulating systems with open boundary conditions, we use a Tensorized version of the GRU cell, which improves the expressiveness of the ansatz~\cite{hibat-allah_variational_2021} at the cost of a larger number of parameters. This cell only accepts the two hidden vectors and spin states that come from the (square-lattice) nearest-neighbor cells that appear previously in the autoregressive sequence. The candidate hidden vector and update gate are computed as follows:
\begin{align*}
    \Tilde{\vec{h}}_{i,j} &= \text{tanh}([\vec{\sigma}_{i-1,j};\vec{\sigma}_{i,j-1}]
    \mathbf{T} [\vec{h}_{i-1,j};\vec{h}_{i,j-1}] + \vec{b}),
    \\
    \vec{u}_{i,j} &= \text{sigmoid}([\vec{\sigma}_{i-1,j};\vec{\sigma}_{i,j-1}]
    \mathbf{T}_g [\vec{h}_{i-1,j};\vec{h}_{i,j-1}] + \vec{b}_g).
\end{align*}

For both types of GRU cell, we compute an output hidden vector from the candidate hidden vector $\Tilde{\vec{h}}_{i,j}$ and the update gate $\vec{u}_{i,j}$,
\begin{align*}
    \vec{h}_{i,j} = \vec{u}_{i,j} \odot\Tilde{\vec{h}}_{i,j} + (1 - \vec{u}_{i,j}) \odot \left([\vec{h}_{i-1,j};\vec{h}_{i,j-1}]
    \mathbf{W}_{\text{merge}}\right).
\end{align*} 
The update gate modulates how different the output hidden vector is from the input hidden vectors.

\section{Training details}
\label{app:training_details}

We start our training procedure by optimizing an RNN wavefunction for the TLAHM with $L=6$, where we can achieve the most accurate results. This training is performed in four stages. In the first stage, we fix the learning rate to $\gamma = 5\times10^{-4}$ and the pseudo-temperature to $T=T_0$. In the second stage, with the learning rate still fixed, we perform variational neural annealing and linearly decrease the pseudo-temperature using
\begin{align*}
    T(t) = T_0 \times \left(1-\frac{t}{N_\text{annealing}}\right),
\end{align*}
where $N_\text{annealing}$ is the number of annealing steps. For each annealing step $t$, we perform $N_\text{equilibrium}$ gradient steps for stability purposes~\cite{hibat-allah_variational_2021}. In this work, we fix $N_\text{equilibrium}=5$ and $N_\text{annealing}=10^4$ so that the total number of training steps in this stage of training is $5\times10^{4}\times s$.
In the third and fourth stages, we decay the learning rate according to the following function
\begin{align*}
    \gamma(t) = \gamma_0\times(1+(t/\delta))^{-1},
\end{align*}
with $\gamma_0 = 5\times10^{-4}$, and $\delta = 5\times10^3\times s$. For all scales $s$, $\gamma(t)\approx5\times10^{-5}$ at the end of the fourth stage of training.
In the fourth and final stage of training for $L=6$, we also apply $C_{6v}$ symmetries and perform symmetry averaging on our wavefunction. Symmetry averaging is only performed in this final stage because it significantly increases the computational cost of training. As discussed in \Cref{app:phase_avg}, we only average the magnitudes of $\psi_\mathcal{W}(\vec\sigma)$ and not the phases. 

As we iteratively retrain our RNN wavefunctions for larger system sizes, we use \Cref{eq:schedule} to control the number of training steps. This parameterized schedule depends on two constants $C$ and $F$, the rate $r$, and the scale $s$. Notably, only the scale $s$ affects the number of training steps for $L=6$. For very large $L$, $N_\text{steps}\approx F$, and we fix $F=2\times10^3$. For all system sizes $L>6$, we fix $C = 101\times10^3$, $\gamma=5\times10^{-5}$, and we continue to apply $C_{6v}$ symmetries. We enforce $U(1)$ symmetry during all stages of the training.

We summarize the training schedule in \Cref{tab:training}.
\begin{table}[h]
    \centering
    \begin{tabular}{c|c|c|c|c|c}
    \hline\hline
        System size & $C$ & $T$ & $\gamma$  & $C_{6v}$ & $U(1)$ \\\hline
        $L=6$ (stage 1) & $1\times10^3$ & $T_0$ & $5\times10^{-4}$ & False & True\\
        $L=6$ (stage 2) & $51\times10^3$ & $T(t)$ & $5\times10^{-4}$ & False & True\\
        $L=6$ (stage 3)  & $76 \times10^3$ & $0$ &  $\gamma(t)$ & False & True\\
        $L=6$ (stage 4)  & $101\times10^3$& $0$ & $\gamma(t)$ & True & True\\
        $L>6$ & $101\times10^3$ & $0$ & $5\times10^{-5}$ & True & True\\
    \hline\hline
    \end{tabular}
    \caption{\small Hyperparameters for the iterative retraining of RNN wavefunctions.}
    \label{tab:training}
\end{table}

\section{Phase averaging\label{app:phase_avg}}

Averaging the phases over the group $\mathcal{G}$ can be done using the following formula~\cite{hibat-allah_recurrent_2020, Reh_optimizing_2023}:
\begin{align}
    \phi'_{\mathcal{W}}(\vec{\sigma}) =  \text{Arg}\left(\sum_{\mathcal{T} \in \mathcal{G}} ^{|\mathcal{G}|} \text{exp}[i\phi'_{\mathcal{W}}(\mathcal{T} \vec{\sigma})]
    \right).\label{eq:phase_avg}
\end{align}

\begin{figure}[]
    \centering
    \includegraphics[width=\linewidth]{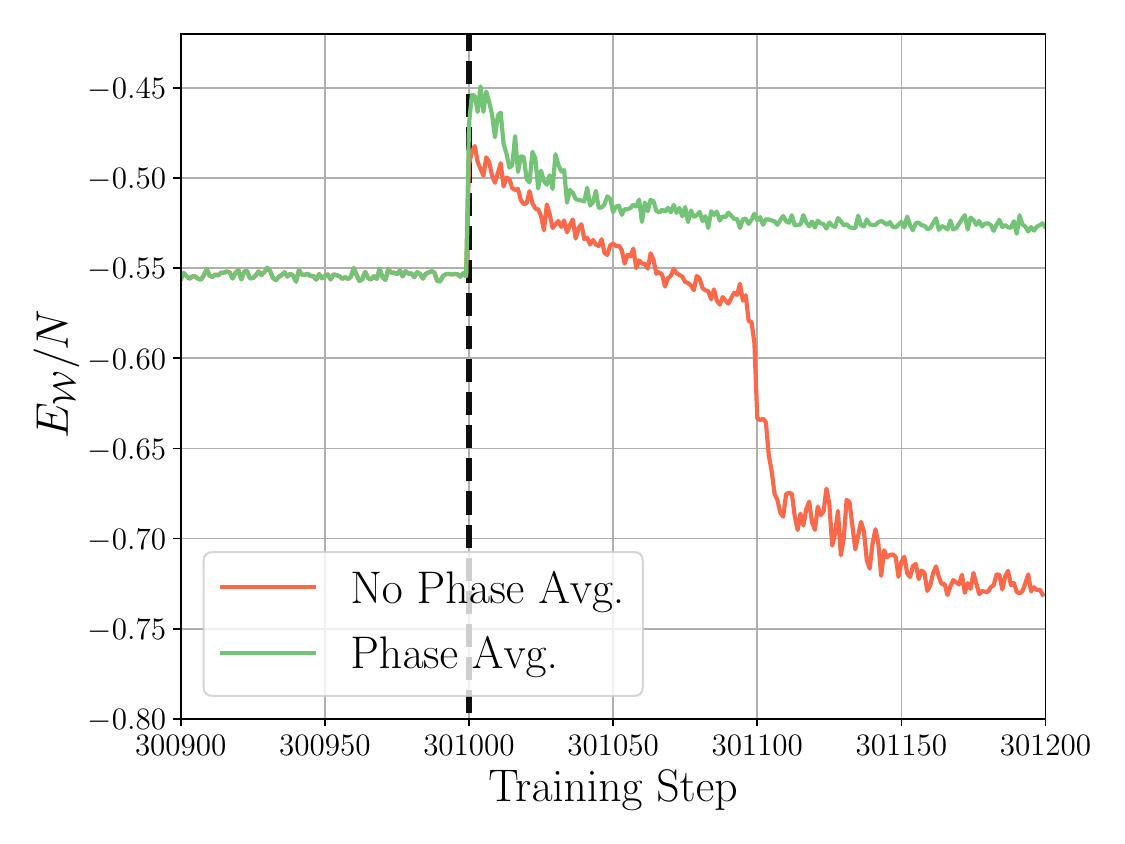}
    \caption{Energy per spin as a function of training step around the point in the training where the lattice symmetries are first enforced (black dashed line). For both optimizations shown here we average the magnitude of the amplitudes according to~\Cref{eq:sym_avg}, but for the green line we also perform the phase averaging according to~\Cref{eq:phase_avg}.}
    \label{fig:phase_average_energies}
\end{figure}

\begin{figure*}[]
    \centering
    \includegraphics[width=\linewidth]{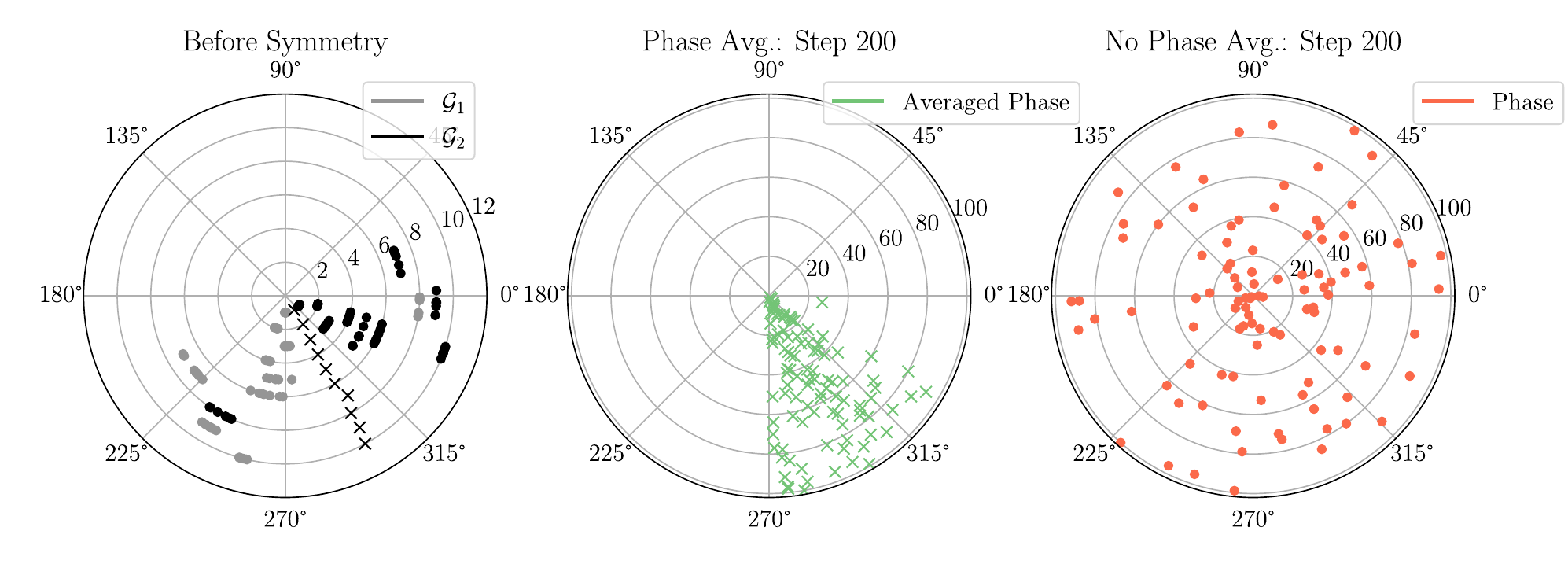}
    \caption{Complex phases of samples drawn from the RNN. The distance from the center of each circle corresponds to the sample index. Points indicate the phases of individual samples, whereas crosses indicate phases obtained by performing the phase average using~\Cref{eq:phase_avg}. We show samples obtained before turning on the symmetries (left), after 200 steps of training with symmetries (center) and without phase averaging (right). (Left) For 10 samples, we show the phases of the $|\mathcal{G}|=12$ symmetry transformed samples $\mathcal{T}\vec{\sigma}$. The phases separate into two sets based on the group actions $\mathcal{G}_1=\{\mathcal{T}(\theta)|\:\theta\in\{0^\circ,120^\circ, 240^\circ\}\}$ and $\mathcal{G}_2=\{\mathcal{T}|\:\theta\in\{60^\circ,180^\circ, 300^\circ\}\}$ where $\mathcal{T}(\theta)=r\circ g(\theta)$ with $r$ a reflection along the diagonal of the lattice and $g(\theta)$ a rotation by an angle of $\theta$. (Center) We show the phases of 100 samples obtained after training with phase averaging. (Right) We show the phases of the original sample $\vec{\sigma}$ which is used during training without phase averaging.}
    \label{fig:phase_average}
\end{figure*}

As observed in Ref.~\cite{Reh_optimizing_2023}, separating the averaging of the magnitude and phase can inhibit the representational power of the network. Here we observe a similar effect in~\Cref{fig:phase_average_energies}, where we plot the variational energies directly after turning on the lattice symmetries in stage 4 of our training (see~\Cref{app:training_details}). With and without phase averaging, the variational energies briefly jump to larger values when the symmetry averaging begins. However, we see that the variational energies remain large when the phases of the amplitudes are averaged, whereas they quickly converge to lower values when the phases are not averaged.

To further investigate the reason behind this observation, we examine the complex phases corresponding to samples drawn from our trained RNN wavefunction before and after we begin enforcing lattice symmetries during training. In \Cref{fig:phase_average},
we see that before the lattice symmetries are enforced during training, when the symmetry transformations $\mathcal{T}$ are applied to a given sample, the phases can be separated into two distinct clusters based on the transformation $\mathcal{T}$ (left panel). The average over the phases of all these symmetrized samples is around $60^\circ$, as shown by the black crosses.
When lattice symmetries are enforced with phase averaging during training, after only 200 training steps we see that the complex phases of our samples start to cluster around $60^\circ$ (center panel). Without phase averaging, however, we observe a much richer sign structure from the samples (right panel), indicating that phase averaging can have adverse effects during training.

\section{Runtime details}
\label{app:runtimes}

The main computational cost during a single training step is the calculation of the variational energy. This is because one must obtain the log amplitudes of all configurations that are connected to the input samples by the Hamiltonian. The number of connected configurations scales as $\mathcal{O}(N)$. For each of these configurations, evaluating the log amplitudes costs $\mathcal{O}(N)$. Therefore, the cost of a single training step should scale as $\mathcal{O}(N^2) = \mathcal{O}(L^4)$~\cite{moss_leveraging_2025}. While the number of parameters in the RNN wavefunction also contributes to the training time, we emphasize that we keep the size of our ans\"{a}tze fixed throughout this work such that our RNN wavefunction is sufficiently expressive and so the scaling behavior is only a function of the system size $L$. \Cref{fig:time_per_L} shows how the time per training step scales in practice and how well our theoretical estimate describes this scaling.

\setcounter{figure}{9}
\begin{figure}[t!]
    \centering
    \includegraphics[width=\linewidth]{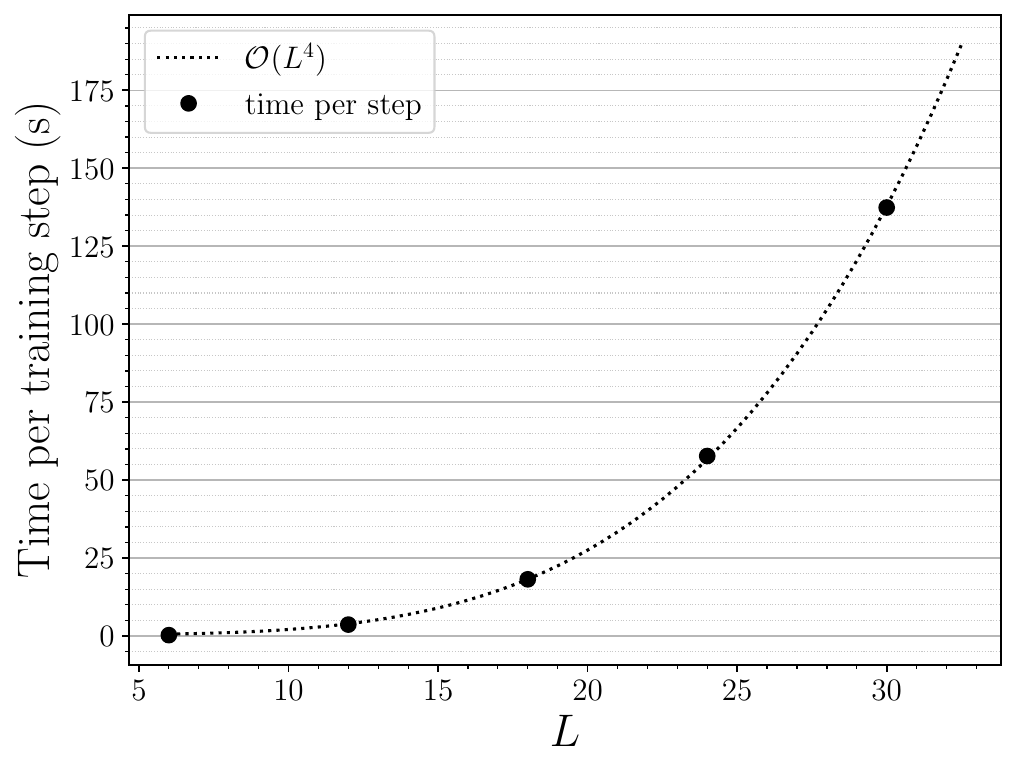}
    \caption{The amount of time in seconds it takes to complete one train step for a given system size with periodic boundary conditions using two H200 GPUs.}
    \label{fig:time_per_L}
\end{figure}

\begin{figure}
    \centering
    \includegraphics[width=\linewidth]{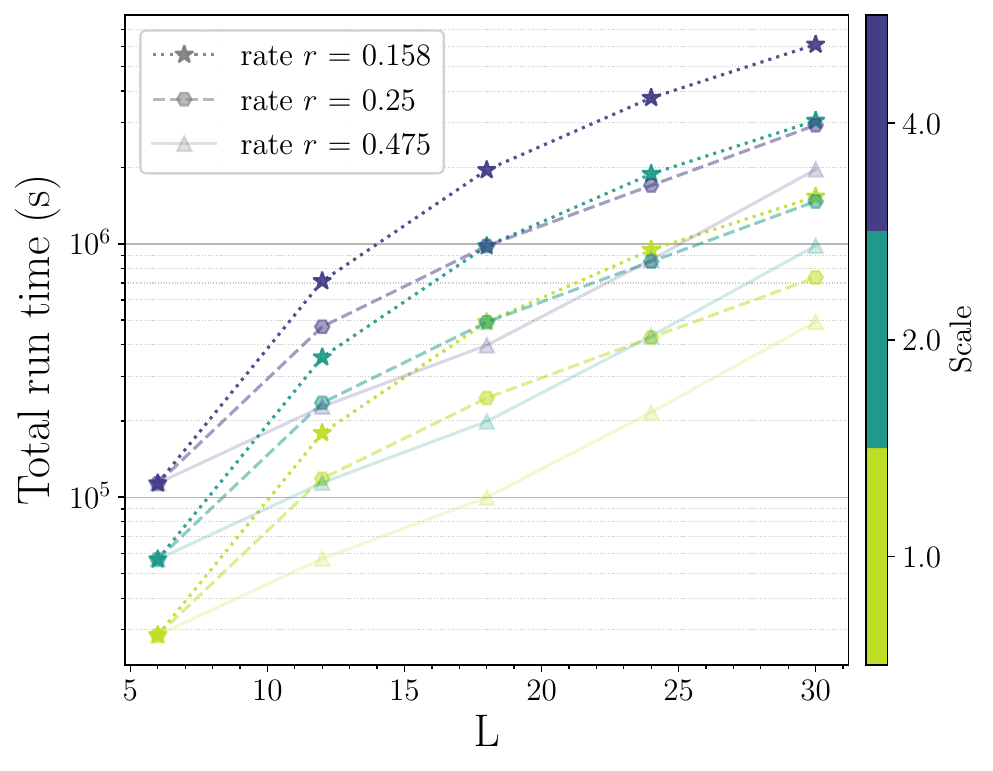}
    \caption{The amount of time in seconds it takes to complete the training up to a given system size with periodic boundary conditions using two H200 GPUs.}
    \label{fig:runtimes}
\end{figure}

As discussed, the parameterized training schedule defined by \Cref{eq:schedule} decreases the number of training steps as the system size increases. In other words, for very large system sizes, where the time per training step is very large, we perform fewer training steps, which prevents the total runtime for our simulations from scaling as rapidly as the time per training step. \Cref{fig:runtimes} shows the total runtime for each training schedule considered in this work. The amount of time shown is cumulative; it represents the total time required to perform training for all lattice sizes up to and including a given $L$. 

We emphasize that this parameterized training schedule would not be sensible if the VMC simulations for every system size $L$ were initialized from scratch. For instance, most NQS architectures do not permit be iterative retraining. Starting from a random initialization should give rise to a more challenging optimization that requires at least the same number of training steps for increasing system size $L$. This is because of the exponential growth of the Hilbert space. Therefore, the ability to iteratively retrain our RNN wavefunction such that we are able to successfully employ our parameterized training schedule significantly reduces the amount of runtime required to obtain the results presented in this work.

\section{V-score scaling}
\label{app:vscores}

Recently, Wu et. al.~\cite{wu_variational_2024} introduced a metric called the V-score, which quantifies the accuracy of the result from a variational method. The V-score is intensive, i.e., independent of system size, and accounts for both the variational energy and the variance of the variational energy, both of which can be used to assess the variational result separately. The V-score is defined as
\begin{align*}
    \text{V-score} := \frac{N\text{var}(E)}{(E-E_\infty)^2},
\end{align*}
where $E_\infty$ is the zero-point energy, which is zero for the Heisenberg model. It has been observed that the V-score scales linearly with the relative error of the variational energy in many cases~\cite{wu_variational_2024,moss_leveraging_2025}, justifying its use.

\begin{figure}[t]
    \centering
    \includegraphics[width=\linewidth]{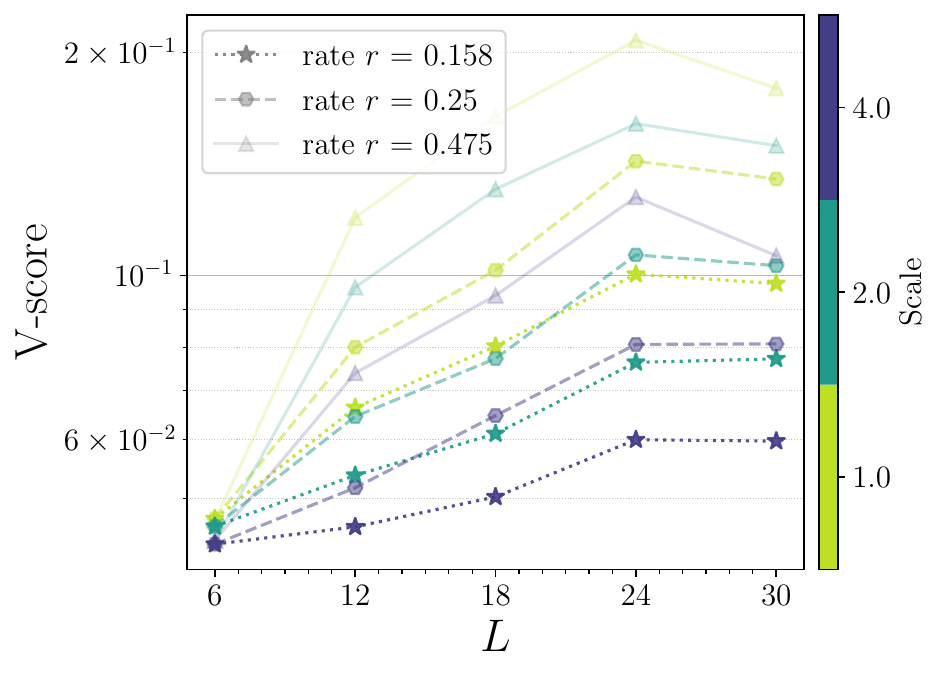}
    \caption{V-scores obtained for all lattice sizes $L$ from all of our RNN wavefunctions. These V-scores are computed using the variational energies and the corresponding variances shown in \Cref{fig:energies_peri}.}
    \label{fig:vscores_peri}
\end{figure}

\Cref{fig:vscores_peri} shows the V-scores corresponding to our results. 
We see that increasing the training time by adjusting the scale $s$ or the rate $r$ leads to a systematic improvement in the V-score, which is consistent with the results presented in the main text. Our V-scores are consistently better than the reference values for the TLAHM reported in Ref.~\cite{wu_variational_2024}. 
We present the V-scores corresponding to the variational results from our most accurate simulations in \Cref{tab:vscores}.

\begin{table}[h]
    \centering
    \begin{tabular}{c|c}
    \hline\hline
    system size $L$ & V-score \\\hline
    6 &  $4.3\times 10^{-2}$ \\ 
    12 & $4.6\times 10^{-2}$ \\ 
    18 & $5.0\times 10^{-2}$ \\ 
    24 & $6.0\times 10^{-2}$ \\ 
    30 & $6.0\times 10^{-2}$ \\ 
    \hline\hline
    \end{tabular}
    \caption{\small V-scores of the variational results from our most accurate simulation of the TLAHM with periodic boundary conditions, scale $s=4.0$ and rate $r=0.158$.}
    \label{tab:vscores}
\end{table}

Part of the motivation behind the V-score is that this metric can be used to compare the quality of variational results independent of the Hamiltonian being studied.
Interestingly, these V-scores are significantly larger than those reported for the square-lattice Heisenberg model in Fig.~10 of Ref.~\cite{moss_leveraging_2025}. This result suggests that learning accurate representations of the ground-states of the TLAHM is a much more challenging task than for the square-lattice Heisenberg antiferromagnet.

\section{Zero-variance extrapolation of the variational energy}
\label{app:zero-var}

\begin{figure}
    \centering
    \includegraphics[width=\linewidth]{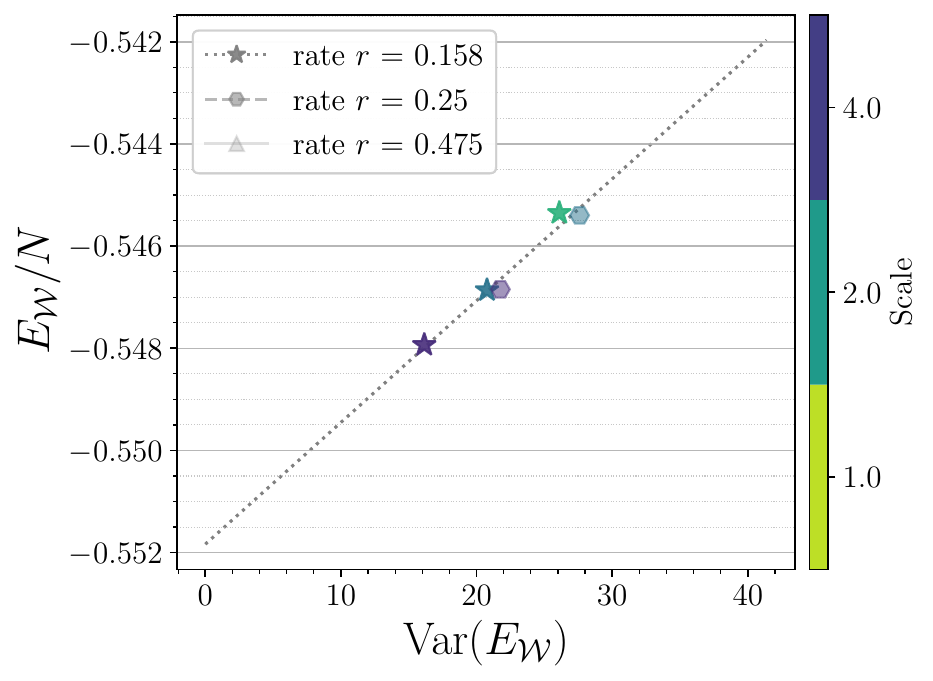}
    \caption{The zero-variance extrapolation of our best variational energies for $L=30$. Each variational energy comes from an RNN wavefunction trained according to the schedule given by \Cref{eq:schedule} and a unique choice of the scale $s$ and rate $r$. The dotted grey line shows the linear fit through the data. }
    \label{fig:zero-var-extrap}
\end{figure}

When one has access to a sequence of systematically improving variational states, it is possible to extrapolate the corresponding variational energies to the zero-variance limit~\cite{nomura_restricted_2017,kwon_effects_1998,becca2000stabilitydwavesuperconductivitytj,Sorella2001,hu_direct_2013,fu_variance_2023}. This type of extrapolation is motivated by the zero-variance principle~\cite{becca_quantum_2017}. \Cref{fig:zero-var-extrap} shows an example of such an extrapolation for our variational energies for $L=30$. As mentioned, we perform this fit only through our best variational energies. 

\section{Estimating the total spin of trained RNN wavefunctions}
\label{app:total_spin}

The total spin $\langle\vec{S}^2\rangle$ can be computed
as
\begin{align*}
    \langle \vec{S}^2\rangle = \frac{3}{4}N + 2\sum_{i<j}C(i,j),
\end{align*}
where $C(i,j)$ is defined by \Cref{eq:real_space_corr}.

Our most accurate RNN wavefunctions learn states with non-zero total spin, supporting the hypothesis that the RNN wavefunctions learn a superposition of the states in Anderson's tower of states. All of the states in the tower of states live in different total spin sectors. 
The values of $\langle\vec{S}^2\rangle$ estimated from our most accurate RNN wavefunctions are summarized in \Cref{tab:total-spin} below. The same RNN wavefunctions produced the results in \Cref{fig:M_scaling_peri}.
\begin{table}[h]
    \centering
    \begin{tabular}{c|c}
        \hline\hline
        $L$ & $\langle\vec{S}^2\rangle$ \\\hline
         6 & 0.0158 \\
         12 & 0.0246 \\
         18 & 0.0259 \\
         24 & 0.0275 \\
         30 & 0.0276 \\
         \hline\hline
    \end{tabular}
    \caption{\small Estimates of the total spin $\langle\vec{S}^2\rangle$ from our most accurate RNN wavefunctions. The wavefunctions are from our simulations with $s=4.0$ and $r=0.158$.}
    \label{tab:total-spin}
\end{table}

To our knowledge, Ref.~\cite{roth_high-accuracy_2023} contains the only other estimates of $\langle\vec{S}^2\rangle$ for the TLAHM obtained with NQS. For $N=36$, Roth \emph{et al.} achieve a much smaller value of $\langle\vec{S}^2\rangle$, and a much more accurate ground state energy. However, for $N=108$, their reported value of $\langle\vec{S}^2\rangle$ is larger than the value we achieve with our RNN wavefunctions for larger system sizes.

\section{Results for the TLAHM with open boundary conditions}
\label{app:obc}

The first attempts to study the TLAHM using iteratively retrained RNN wavefunctions only examined systems with open boundary conditions~\cite{hibat-allah_supplementing_2024}. Furthermore, those results were obtained by transforming the TLAHM Hamiltonian according to the local unitary transformation given by the Marshall-Peierls sign rule, $\mathcal{U}_\text{sq}$ defined in \Cref{eq:sq_sign_rule} and by considering systems with even lattice lengths. Not all lattices with even lattice length are commensurate with the three-sublattice ordering of the ground state of the TLAHM.
For completeness, we reproduce those results, but we also expand on them. In particular, we study how the results change if we perform iterative retraining using only commensurate lattice sizes and if we use the $120^\circ$ transformation, $\mathcal{U}_\text{tri}$ defined in \Cref{eq:tri_sign_rule}, instead of $\mathcal{U}_\text{sq}$. We emphasize that the commensurate lattice sizes we consider are lattices with lengths $L$ that are multiples of 6.
All of the following results are based on three different simulations, which are summarized in \Cref{tab:obc_experiments}. We follow the same training procedure outlined in \Cref{app:training_details}.

\begin{table}[h]
    \centering
    \begin{tabular}{c|c|c|c}
    \hline\hline
        basis transformation & lattice sizes & scale & rate \\
        applied to $\hat{H}$ & $L$ $(n \in \mathbb{Z}^+)$ & $s$ & $r$ \\
        \hline
        $\mathcal{U}_\text{sq}$ & $2n$ & 1.0 & 0.475 \\
        $\mathcal{U}_\text{sq}$ & $6n$ & 1.0 & 0.158\\
        $\mathcal{U}_\text{tri}$ & $6n$ & 1.0 & 0.158\\
    \hline\hline
    \end{tabular}
    \caption{The specifications for the three simulations performed for the TLAHM with open boundary conditions. When $s=1.0$ and $r=0.475$, our training schedule, obtained using \Cref{eq:schedule}, is very close to what was used in Ref.~\cite{hibat-allah_supplementing_2024}. When only commensurate lattices sizes are studied, we reduce the rate to $r=0.158$ which is roughly 3 times smaller than $r=0.475$ so that the RNN wavefunction is optimized for the same number of training steps for the second lattice in the iterative retraining procedure, whether that lattice length is $L=8$ or $L=12$.}
    \label{tab:obc_experiments}
\end{table}

\begin{figure}
    \centering
    \includegraphics[width=\columnwidth]{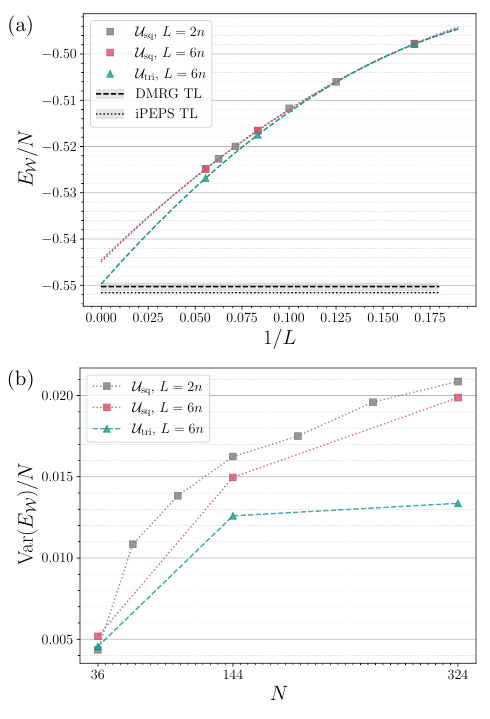}
    \caption{(a) The variational energies for all system sizes obtained from each simulation described by \Cref{tab:obc_experiments}. These energies are scaled according to $1/L$ and fit with a second-order polynomial. The reference values of the ground-state energy in the thermodynamic limit (TL) are shown for comparison. The dashed line is the value from DMRG simulations using MPS with cylindrical boundary conditions~\cite{huang_magnetization_2024}. The dotted line is the value from variationally-optimized iPEPS~\cite{hasik_incommensurate_2024}. Each variational energy is estimated with $10\times10^3$ samples.
    (b) The variances of the final variational energies shown above, plotted as a function of the number of spins in the system $N$.}
    \label{fig:energies_open}
\end{figure}

\Cref{fig:energies_open}(a) shows the variational energies obtained from each of these simulations scaled as $1/L$. We are able to extract estimates of the ground-state energy in the thermodynamic limit by fitting a second-order polynomial in $1/L$ to these finite-size variational energies. Notably, we perform this fit directly with the variational energies from each simulation. In other words, we do not perform a zero-variance extrapolation, since we perform only one simulation with $s=1.0$ and $r=0.475$. We see from the finite-size energies and the extrapolations to the thermodynamic limit that the $120^\circ$ transformation significantly improves our results. In fact, our scaling of the variational energies from the simulation where the $120^\circ$ transformation was used yields a very accurate estimate of the ground-state energy in the thermodynamic limit $E_\infty = -0.5497$, which is within the error bars of the reference DMRG value $E_\infty^\text{DMRG} = -0.5503(8)$. A simulation with a larger scale or a slower rate would likely yield improved variational energies such that the extrapolated ground-state energy is closer to the reference iPEPS value. 
In \Cref{fig:energies_open}(b), we show the variances of the variational energies displayed in (a). In accordance with the energies, the simulation where we employed the $120^\circ$ transformation produces the lowest variances. 

\begin{figure}
    \centering
    \includegraphics[width=\columnwidth]{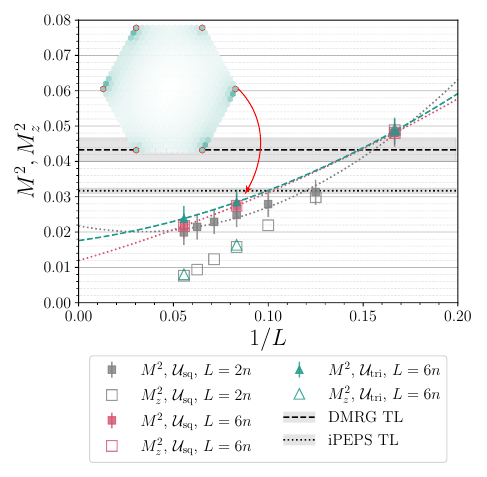}
    \caption{Our estimates of the squared sublattice magnetization scaled as a function of $1/L$. We show estimates of $M^2$ and $M_z^2$. The closed markers correspond to the values of $M^2$ estimated using the correlations defined in \Cref{eq:real_space_corr} and open markers correspond to the values of $M^2_z$, which are estimated using the correlations defined by \Cref{eq:real_space_corr_z}.  The reference values for the squared sublattice magnetization in the thermodynamic limit (TL) are shown for comparison. The dashed line is the value from DMRG simulations using MPS with cylindrical boundary conditions~\cite{huang_magnetization_2024}. The dotted line is the value from variationally-optimized iPEPS~\cite{hasik_incommensurate_2024}. 
    The inset shows all of the momentum-space correlations captured by our RNN wavefunction for $L=12$ when $\mathcal{U}_\text{tri}$ is employed. The points in momentum space corresponding to the expected ordering wavevectors are highlighted in red.}
    \label{fig:M_scaling_open}
\end{figure}

Despite the accuracy of the variational energies, it is important to assess whether our simulations have accurately captured other important physical quantities of the target ground states. We also estimated the sublattice magnetization using our trained RNN wavefunctions. \Cref{fig:M_scaling_open} shows finite-size estimates of the squared sublattice magnetization calculated according to \Cref{eq:M}. We show the values of this quantity estimated with the correlations defined by \Cref{eq:real_space_corr} and the correlations defined by \Cref{eq:real_space_corr_z}. We also fit a second-order polynomial to these estimates in order to estimate the value of the sublattice magnetization in the thermodynamic limit. 

Interestingly, for these simulations, we do not always observe that the RNN wavefunctions break the SU(2) symmetry of the finite-size ground states. For $L=6$, it appears that all the learned ground states are SU(2) symmetric, as seen by the agreement between $M^2$ and $M^2_z$ in \Cref{fig:M_scaling_open}. When the RNN wavefunctions do break the symmetry, it is less dramatic than what we observe from our simulations of systems with periodic boundary conditions; see \Cref{fig:M_scaling_peri}. It is possible that additional training time, e.g., a larger scale $s$ or slower rate $r$, would allow the wavefunction to maintain a symmetric ground state throughout the iterative retraining process. 

\begin{figure}
    \centering
    \includegraphics[width=\columnwidth]{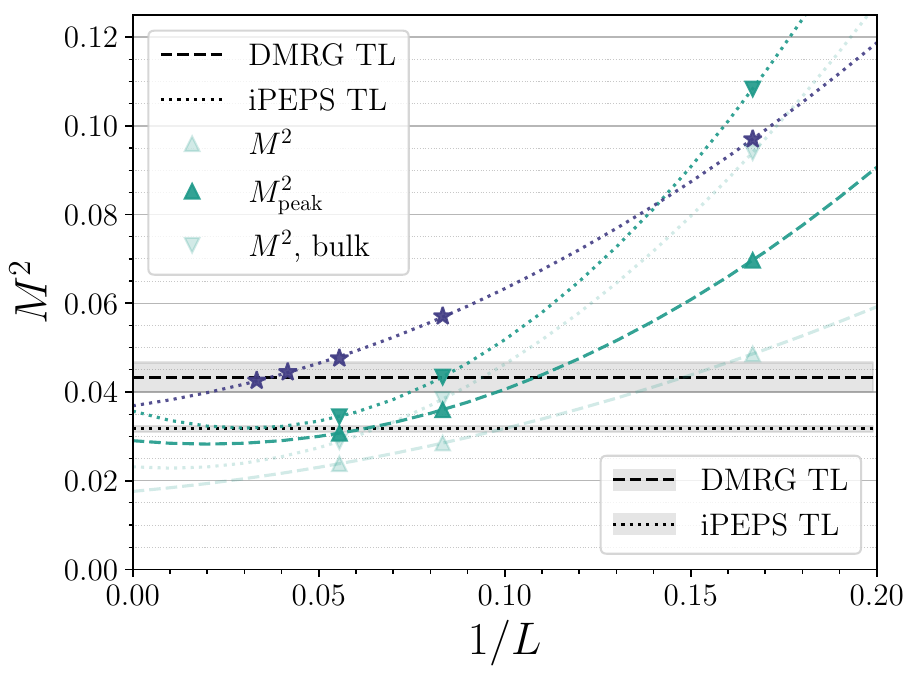}
    \caption{Our estimates of the squared sublattice magnetization scaled as a function of $1/L$. These estimates come from our simulation where we employed $\mathcal{U}_\text{tri}$ and considered only commensurate lattice sizes $L=6n$. Here we consider many different estimates of $M^2$. We show \Cref{eq:M} evaluated at the k-point corresponding to the expected ordering wavevector (shown in \Cref{fig:M_scaling_open}) and evaluated using the actual peak value of the momentum-space correlations. Furthermore, we estimate $M^2$ using only the correlations in the bulk of our system, again evaluating \Cref{eq:M} at the k-point corresponding to the expected ordering wavevector and using the actual peak value of the momentum-space correlations. The reference values for the squared sublattice magnetization in the thermodynamic limit (TL) are shown for comparison. The dashed line is the value from DMRG simulations using MPS with cylindrical boundary conditions~\cite{huang_magnetization_2024}. The dotted line is the value from variationally-optimized iPEPS~\cite{hasik_incommensurate_2024}. }
    \label{fig:M_scaling_bulk_open}
\end{figure}

The estimates of the squared sublattice magnetization, for both finite sizes and in the thermodynamic limit, are significantly reduced from what we obtain from our simulations for systems with periodic boundary conditions and from the reference values. We believe this result is due to strong boundary effects, which can significantly corrupt the order in the ground state, even for systems with $L=18$, which is the largest lattice size considered in these simulations. The full picture of the momentum-space correlations, shown in the inset of \Cref{fig:M_scaling_open}, exhibit the corruption of the magnetic order. The peaks of the momentum-space correlations are not located at the points corresponding to the expected ordering wavevectors, $\vec{q} = (\frac{4\pi}{3}, 0)$ and its rotations. Instead, they are slightly shifted. Furthermore, the magnitude of these peaks is reduced from what we observed from our simulations of the TLAHM Heisenberg model with periodic boundary conditions. This observation is illustrated in \Cref{fig:M_scaling_bulk_open}, where we examine the estimates of the squared sublattice magnetization from our simulation using $\mathcal{U}_\text{tri}$ more carefully. We show the values of $M^2_\text{peak}$, which is calculated according to 
\begin{align*}
    M^2_\text{peak} = \frac{\text{max}(S(L,\vec{q}))}{L^2}.
\end{align*}
While our estimates of $M^2_\text{peak}$ are larger than our estimates of $M^2$, they are still much lower than the values of $M^2$ obtained from our simulations for systems with periodic boundary conditions, and they extrapolate to a value that is still considerably smaller than the reference value.

When performing finite-size scaling studies for systems with open boundary conditions, another method for reducing boundary effects is to consider only the correlations in the bulk of the system~\cite{liu_gradient_2017}. \Cref{fig:M_scaling_bulk_open} also shows estimates of $M^2$ and $M^2_\text{peak}$ when they are calculated using only correlations from the bulk of the system.
Even then, we are unable to significantly improve our estimates of the squared sublattice magnetization.

These results give a clear example of why it is important to examine correlations in addition to variational energies. In general, for systems with open boundary conditions, it is easier to obtain good variational results, as seen through the variational energies, the variances of those energies, or even the V-scores~\cite{wu_variational_2024}, but boundary effects can preclude one's ability to extract other important quantities that are functions of the correlations in the system in the thermodynamic limit through a finite-size scaling study.

\onecolumngrid
\section{Comparison of results to benchmarks and other methods}
\label{app:comparison}

In this appendix we directly compare our results to the existing literature. To our knowledge, there is no published data set for the TLAHM on finite-size lattices that includes all of the lattice sizes we consider in this work, meaning we cannot carefully benchmark every one of our finite-size results. However, a common benchmark is the $L=6$ triangular lattice with periodic boundary conditions. In \Cref{tab:L=6}, we compare our ground-state energy per spin for the $L=6$ TLAHM to other variational results. We also benchmark our estimates of the thermodynamic limit properties of the ground state with other available reference values. This comparison can be found in \Cref{tab:TL}.

\begin{table*}[h!]
    \centering
    \begin{tabularx}{\textwidth}{X X X}
        \hline\hline
        Ansatz and reference & & $E/N$   \\
        \hline
        Projected mean field~\cite{weber_magnetism_2006} & & $-0.543(1)$ \\
        Projected ansatz~\cite{iqbal_spin_2016} & & $-0.548025(3)$ \\
        Projected mean field ansatz~\cite{heidarian_spin-frac12_2009} & & $-0.55148(5)$ \\
        Restricted Boltzmann Machine~\cite{ferrari_neural_2019} & & $-0.553$ \\
        Entangled-plaquette state~\cite{mezzacapo_ground-state_2010} & & $-0.55420(5)$ \\
        Projected mean field ansatz~\cite{kaneko_gapless_2021} & & $-0.55519(4)$ \\
        \bf{RNN wavefunction} & & $\mathbf{-0.5562(2)}$ \\
        Group Convolutional Neural Network~\cite{roth_group_2021} & & $-0.55922$ \\
        Lattice Convolutional Network~\cite{fu_lattice_2022} & & $-0.5601(4)$ \\
        Group Convolutional Neural Network~\cite{roth_high-accuracy_2023}& & $-0.560313(3)$ \\
        \hline
        Exact Diagonalization~\cite{bernu_exact_1994} & & $-0.5603734$ \\
        \hline\hline
    \end{tabularx}
    \caption{Estimates of the ground-state energy per spin for the TLAHM on an $L=6$ lattice with periodic boundary conditions. This list summarizes variational results from the references discussed in the introduction, which considered the same geometry. Our results, which are shown in bold in this table, are those obtained from our most accurate simulation (scale $s=4.0$ and rate $r=0.158$). We show the exact value of this energy, obtained with exact diagonalization~\cite{bernu_exact_1994}, for reference. }
    \label{tab:L=6}
\end{table*}

\begin{table*}[h!]
    \centering
    \begin{tabularx}{\textwidth}{X X X}
        \hline\hline
        Method and reference & $E_\infty/N$ & $M_\infty$  \\
        \hline
        Spin-Wave Theory~\cite{jolicoeur_spin-wave_1989} & $-0.5388$ & 0.239 \\
        Spin-Wave Theory~\cite{j_miyake_spin-wave_1992} & $-0.5466$ & 0.2497\\
        \hline 
        Exact Diagonalization~\cite{bernu_exact_1994} & $-0.5445$ & -- \\      
        Exact Diagonalization~\cite{bernu_signature_1992} & $-0.5475$ & 0.25 \\ 
        \hline
        Series expansion~\cite{zheng_excitation_2006} & 
        $-0.5502(4)$ & 0.19(2)\\
        \hline
        QMC - Green's function + SR~\cite{capriotti_long-range_1999} & $-0.5458(1)$ & 0.205(10) \\
        \hline
        VMC - RVB ansatz~\cite{anderson_resonating_1973,fazekas_ground_1974} & $-0.463(7)$, $-0.54(1)$ & -- \\
        VMC - projected mean field ansatz~\cite{weber_magnetism_2006} & $-0.532(1)$ & $\sim$0.36  \\
        VMC - Jastrow ansatz~\cite{huse_simple_1988} & $-0.5367$ & $\sim$0.34 \\
        VMC - projected mean field ansatz~\cite{kaneko_gapless_2021} & $-0.5449(2)$ & 0.271(3) \\
        VMC + Lanczos - projected ansatz~\cite{iqbal_spin_2016} & $-0.545321(7)$ & -- \\
        VMC - projected mean field ansatz~\cite{heidarian_spin-frac12_2009} & $-0.5470(1)$ & 0.264 \\
        \bf{VMC - RNN (this work)} & $\mathbf{-0.5497}$  (OBC), & -- \\
        & $\mathbf{-0.5517569(9)}$  (PBC) & $\mathbf{0.192(2)}$ ($M^2$), $\mathbf{0.198(2)}$ ($M^2_C$) \\
        \hline 
        two-dimensional DMRG~\cite{xiang_two-dimensional_2001} & $-0.5442$ & -- \\
        DMRG (cylinders)~\cite{white_neel_2007} & -- & 0.205(15)\\
        DMRG (cylinders)~\cite{huang_magnetization_2024} & $-0.5503(8)$ & 0.208(8)\\
        iPEPS~\cite{hasik_incommensurate_2024} & $-0.55161(6)$ & $0.178(2)$, $0.159(6)$\\
        \hline\hline
    \end{tabularx}
    \caption{Values for the energy per spin and the sublattice magnetization of the TLAHM in the thermodynamic limit from the literature. This list is not comprehensive; it summarizes some of the results discussed in the introduction and includes the state-of-the-art. Our results are shown in bold.}
    \label{tab:TL}
\end{table*}
\twocolumngrid

\clearpage
\bibliography{main}
\end{document}